\documentclass[twocolappendix,11pt,a4paper]{emulateapj}

\usepackage{psfrag}
\usepackage{psfig}
\usepackage{pspicture}
\usepackage{graphicx}

\usepackage{amssymb,amsmath}
\usepackage{natbib}
\def\apj{ApJ}

\submitted{Accepted for publication in the \apj}

\begin{document}

\title {Quenching or Bursting: the Role of Stellar Mass, Environment, and Specific Star Formation Rate to \lowercase{$z\sim$} 1}

\author{
Behnam Darvish\altaffilmark{1}, 
Christopher Martin\altaffilmark{1},
Thiago S. Gon\c{c}alves\altaffilmark{2},
Bahram Mobasher\altaffilmark{3},
Nick Z. Scoville\altaffilmark{1},
and David Sobral\altaffilmark{4,5}
}

\setcounter{footnote}{0}
\altaffiltext{1}{Cahill Center for Astrophysics, California Institute of Technology, 1216 East California Boulevard, Pasadena, CA 91125, USA; email: bdarv@caltech.edu}
\altaffiltext{2}{Observatorio do Valongo, Universidade Federal do Rio de Janeiro, Ladeira Pedro Antonio, 43, Saude, Rio de Janeiro-RJ 20080-090, Brazil}
\altaffiltext{3}{University of California, Riverside, 900 University Ave, Riverside, CA 92521, USA}
\altaffiltext{4}{Department of Physics, Lancaster University, Lancaster, LA1 4YB, UK}
\altaffiltext{5}{Leiden Observatory, Leiden University, P.O. Box 9513, NL-2300 RA Leiden, The Netherlands}

\begin{abstract}

Using a novel approach, we study the quenching and bursting of galaxies as a function of stellar mass ($M_{*}$), local environment ($\Sigma$), and specific star-formation rate (sSFR) using a large spectroscopic sample of $\sim$ 123,000 $GALEX$/SDSS and $\sim$ 420 $GALEX$/COSMOS/LEGA-C galaxies to $z$ $\sim$ 1. We show that out to $z$ $\sim$ 1 and at fixed sSFR and local density, on average, less massive galaxies are quenching, whereas more massive systems are bursting, with a quenching/bursting transition at log($M_{*}$/$M_{\odot}$) $\sim$ 10.5-11 and likely a short quenching/bursting timescale ($\lesssim$ 300 Myr). We find that much of the bursting of star-formation happens in massive (log($M_{*}$/$M_{\odot}$) $\gtrsim$ 11), high sSFR galaxies (log(sSFR/Gyr$^{-1}$) $\gtrsim$ -2), particularly those in the field (log($\Sigma$/Mpc$^{-2}$) $\lesssim$ 0; and among group galaxies, satellites more than centrals). Most of the quenching of star-formation happens in low-mass (log($M_{*}$/$M_{\odot}$) $\lesssim$ 9), low sSFR galaxies (log(sSFR/Gyr$^{-1}$) $\lesssim$ -2), in particular those located in dense environments (log($\Sigma$/Mpc$^{-2}$) $\gtrsim$ 1), indicating the combined effects of $M_{*}$ and $\Sigma$ in quenching/bursting of galaxies since $z$ $\sim$ 1. However, we find that stellar mass has stronger effects than environment on recent quenching/bursting of galaxies to $z$ $\sim$ 1. At any given $M_{*}$, sSFR, and environment, centrals are quenchier (quenching faster) than satellites in an average sense. We also find evidence for the strength of mass and environmental quenching being stronger at higher redshift. Our preliminary results have potential implications for the physics of quenching/bursting in galaxies across cosmic time.     

\end{abstract}
     

\section{Introduction} \label{intro1}

What causes galaxies to stop forming stars --- to quench --- is still an unsolved problem in studies of galaxy formation and evolution. Several external and internal mechanisms with different quenching timescales have been proposed such as ram pressure stripping, viscous stripping, thermal evaporation, strangulation, galaxy-galaxy interactions, galaxy harassment, mergers, galaxy-cluster tidal interactions (see the review by \citealp{Boselli06}), halo quenching \citep{Birnboim03},  AGN feedback (see the review by \citealp{Fabian12}), stellar feedback \citep{Hopkins14}, and morphological quenching and secular processes \citep{Sheth05,Martig09,Fang13,Bluck14,Cavalcante18}.

These processes might temporarily enhance star-formation in galaxies prior to quenching, or they can cause both negative (quenching) and positive (bursting) feedback. For example, compression of the gas due to thermal instability and turbulent motions and/or the inflow of gas to the center can elevate star-formation in galaxies being stripped as a result of ram pressure, prior to the full interstellar medium (ISM) removal of galaxies and hence subsequent quenching \citep{Bekki03,Poggianti16,Poggianti17}. Galaxy-galaxy interactions might cause the gas in the periphery of the interacting systems to get compressed and funnel towards the center, triggering a starburst and/or reviving nuclear activity \citep{Mihos92,Mihos96,Kewley06,Ellison08,Ellison13,Sobral15,Stroe15}. AGN feedback can both reduce/stop star-formation through quasar- and radio-mode feedback \citep{Best05,Croton06,Somerville08,Hopkins10,Gurkan15} and also trigger star-formation by compressing gas (by generating cool, dense cavities in the cocoon around the AGN jet; see e.g.; \citealp{Silk10,Gaibler12,Wagner12,Kalfountzou17}).

More importantly, one particular concern in the studies of galaxy evolution is the assumption that galaxies migrate from the blue cloud to the red sequence (i.e.; they quench) gradually or quickly, whereas in principle, they can also burst and rejuvenate as they evolve. For example, using a new method that makes no prior assumption about the star-formation history of galaxies, \cite{Martin17} show that in-transition green valley galaxies in the local-universe are both quenching and bursting, although the overall mass flux from the blue cloud to the red sequence is positive (quenching). Therefore, to have a better picture of galaxy formation and evolution, we need to simultaneously study and quantify both the ``quenching'' and ``bursting'' of galaxies.  

These processes are directly or indirectly associated with the ``environment'' or ``stellar mass'' of galaxies and they often work together in the quenching mechanism \citep{Peng10,Quadri12,Lee15,Darvish16,Henriques17,Nantais17,Kawinwanichakij17,Guo17,Smethurst17}. The general picture is that the ``environmental quenching'' becomes important at later times (e.g.; \citealp{Peng10,Darvish16,Hatfield17}), particularly for less-massive galaxies \citep{Peng10,Quadri12,Lee15} and ``mass quenching'' is more effective on more massive galaxies especially at higher redshifts \citep{Peng10,Lee15,Darvish16}. In groups, the environmental quenching is thought to be mostly associated with satellites, whereas mass quenching is mainly linked to centrals \citep{Peng12,Kovac14,Darvish17}. However, there are also inconsistencies in the literature on this topic. For example, although some studies point toward an independence of mass quenching and environmental quenching processes \citep{Peng10,Quadri12,Kovac14}, others find that they depend on each other \citep{Darvish16,Kawinwanichakij17}. Despite recent progress, the relative importance of environmental and mass quenching, their evolution with cosmic time, and their influence on the physical properties of galaxies are still not fully understood.

In addition to stellar mass and the environment, another parameter that is strongly linked to galaxy quenching is the specific star-formation rate (sSFR; SFR/$M_{*}$). The inverse of sSFR is a measure of how long it takes a galaxy to assemble its mass given its current SFR. Therefore, it is used to separate star-forming and quiescent systems with the separating sSFR of $\approx$ 10$^{-1}$-10$^{-2}$ Gyr$^{-1}$. The sSFR is tightly coupled to $M_{*}$ for both star-forming and quiescent systems over a broad redshift range \citep{Noeske07a,Wuyts11,Whitaker12,Speagle14,Shivaei15b}. The sSFR also depends on the environment and on average, it is lower in denser regions, particularly at lower redshifts \citep{Peng10,Sobral11,Scoville13,Darvish16,Hatfield17}. However, the cause of lower sSFR in denser environments is still debatable, with some studies attributing this to only a lower fraction of star-forming galaxies in denser regions \citep{Patel09,Peng10,Koyama13a,Darvish14,Darvish15b,Darvish16,Hung16,Duivenvoorden16,Berti17}, whereas others linking it to both a lower fraction and a lower SFR of star-forming galaxies in denser environments than the field \citep{Vulcani10,Patel11,Haines13,Erfanianfar16,Darvish17}. Nonetheless, the latter studies often find a small reduction of $\sim$ 0.1-0.3 dex in star-formation activity of star-forming galaxies in denser regions.  
       
In this paper, we investigate both ``quenching'' and ``bursting'' of the overall galaxy population, satellite galaxies and centrals as a function of four main parameters: stellar mass, sSFR, local environment, and redshift since $z$ $\sim$ 1, based on the recent methodology developed by \cite{Martin17}. In Section \ref{data}, we introduce the data. Methods used to quantify the environment, quenching/bursting of galaxies and their properties are developed in Section \ref{method}. The results are presented in Section \ref{science}, discussed in Section \ref{dis}, and summarized in Section \ref{sum}.           

Throughout this study, we assume a flat $\Lambda$CDM cosmology with $H_{0}$=70 km s$^{-1}$ Mpc$^{-1}$, $\Omega_{m}$=0.3, and $\Omega_{\Lambda}$=0.7 and a Salpeter initial mass function (IMF; \citealp{Salpeter55}). As presented in Section \ref{property}, we define the Star Formation Acceleration (SFA) in units of mag Gyr$^{-1}$ as $\frac{d(NUV-i)_{0}}{dt}$ where $dt$ is the past 300 Myr and $(NUV-i)_{0}$ is the extinction-corrected $NUV-i$ color and the Star Formation Jerk (SFJ) as $\frac{d(NUV-i)_{0}}{dt}$ where $dt$ is the past 600-300 Myr. A positive (negative) SFA and SFJ indicate recent quenching (bursting). The SFA (SFJ) uncertainties are estimated as $\sigma$/$\sqrt{N}$, where $\sigma$ is 1.4826 $\times$ the median absolute deviation of the SFA (SFJ) and $N$ is the number of data points.

\section{Data and Sample Selection} \label{data}

\subsection{Local Universe Sample (SDSS)} \label{datasdss}

\begin{figure*}
\centering
\includegraphics[width=7.0in]{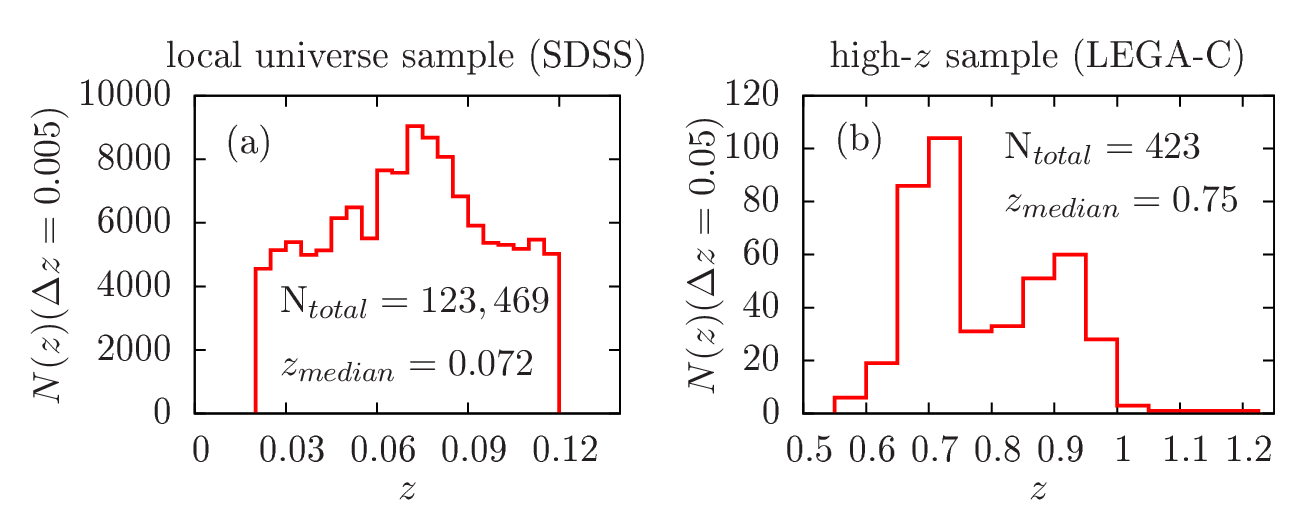}
\caption{(a) Spectroscopic redshift distribution (in bins of $\Delta z$=0.005) of our local-universe SDSS sample. (b) Spectroscopic redshift distribution (in bins of $\Delta z$=0.05) of our high-$z$ LEGA-C sample.}
\label{fig:z-dist}
\end{figure*}

The local universe data are from the SDSS DR12 \citep{Alam15}. Following \cite{Baldry06}, we select galaxies with clean Galactic-extinction-corrected Petrosian magnitude of $r$ $\leqslant$ 17.7 (after excluding stars), clean spectra (after removing duplicates) in the spectroscopic redshift range of 0.02 $\leqslant$ $z$ $\leqslant$ 0.12, located in the contiguous northern galactic cap (130.0 $\leqslant$ RA (deg) $\leqslant$ 240.0 and 0.0 $\leqslant$ Dec (deg) $\leqslant$ 60.0). We use this sample (Sample A) for environmental measure estimations as it provides a contiguous field with relatively uniform, large spectroscopic coverage and completeness. Our estimation of galaxy properties requires SDSS and \textit{GALEX} photometry \citep{Martin05}, 4000 \AA \: break (D$_{n}$(4000)) and H$\delta$ absorption-line index \footnote{The role of SDSS limited fiber size (3$\arcsec$) has been discussed in \cite{Martin07,Martin17}. \cite{Martin17} found no significant effect on their results. As a sanity check, we also limit our sample to $z$=0.04-0.12 and find that our results still hold.} (see Section \ref{property}). Therefore, we match sample A with the \textit{GALEX} All-Sky Survey Source Catalog (GASC; \citealp{Seibert12}) (matching radius of 5$\arcsec$) and the resulting catalog is later matched with the MPA-JHU DR8 catalog \citep{Kauffmann03} to retrieve reliable D$_{n}$(4000) and H$\delta$ (median signal-to-noise (S/N) per pixel $>$ 3) The k-correction recipe of \cite{Chilingarian10} and \cite{Chilingarian12} is used to estimate the rest-frame colors and magnitudes. The final sample comprises 123,469 sources. Figure \ref{fig:z-dist} (a) shows the redshift distribution of sources. We use this final local-universe sample for scientific analysis (Section \ref{science}).

The magnitude cut of $r$ $\leqslant$ 17.7 results in a redshift-dependent stellar mass completeness limit. We estimate the mass completeness limit using \cite{Pozzetti10}. We assign a limiting mass to each galaxy that corresponds to the stellar mass the galaxy would have if its apparent magnitude were the same as the magnitude limit of the sample ($r$ $\leqslant$ 17.7) At each redshift, the 90\% mass completeness, for instance, is then defined as the stellar mass for which 90\% of galaxies have their limiting mass below it. We use this 90\% cut and estimate the completeness limit to be log($M_{*}^{comp}/M_{\odot}$) $\sim$ 10.3 to $z$=0.12.

\subsection{High Redshift Sample (LEGA-C)}

As we already mentioned, we require high signal-to-noise D$_{n}$(4000) and H$\delta$ absorption features (along with photometric information) to robustly extract galaxy properties. At higher redshifts, the only such large and deep galaxy sample available so far is from the VLT LEGA-C spectroscopic survey \citep{vanderwel16} in the COSMOS field \citep{Scoville07} at $z$ $\approx$ 0.6-1.0. Similar to the SDSS quality, this survey is designed to obtain high resolution ($R$ $\sim$ 2500), high S/N ($\gtrsim$ 10, through 20 hour integration) continuum spectra in the wavelength range of $\sim$ 6300-8800 \AA \: for a large ($\sim$ 3200) sample of galaxies at $z$ $\sim$ 1 using the VIMOS spectrograph. Their primary sample is $K$-band selected with a redshift-dependent magnitude limit to guarantee the coverage of the full galaxy types including quiescent, star-forming, and dusty systems at log($M_{*}$/$M_{\odot}$) $\gtrsim$ 10 (Chabrier IMF).

We use the LEGA-C first data release (892 spectra) by selecting galaxies with continuum S/N $>$ 3 (typical S/N $>$ 10) and available D$_{n}$(4000) and H$\delta$ indices\footnote{For both the SDSS and LEGA-C samples, we use the definition of \cite{Balogh99} in order to extract D$_{n}$(4000) and H$\delta$.}. We match this sample with the $i^{+}$-band selected catalog of \cite{Capak07} to obtain GALEX $FUV$/$NUV$ \citep{Zamojski07}, $CFHT$ $u^{*}$, and $Subaru$ $g^{+}$, $r^{+}$, and $i^{+}$ photometry. We convert the $CFHT$ $u^{*}$ magnitude to the SDSS using $u^{*}$=$u$-0.241($u$-$g$) (from the $CFHT$ website). $Subaru$ $g^{+}$, $r^{+}$, and $i^{+}$ magnitudes are converted to SDSS using table 8 in \cite{Capak07}. k-correction is evaluated using the best-fit SED template at the redshift of the sources \citep{Ilbert09}. The final sample contains 423 galaxies, spanning 0.6 $\lesssim$ $z$ $\lesssim$ 1.0 (median redshift of $z_{median}$ $\approx$ 0.75), with the mass completeness limit of log($M_{*}^{comp}/M_{\odot}$) $\sim$ 10.3 \citep{vanderwel16}. Figure \ref{fig:z-dist} (b) shows the redshift distribution of our high-$z$ sample.  
            
\section{Methods} \label{method}

\subsection{Local Environment}

There are different measures for defining the ``environment'' of galaxy on different physical scales, with each method having its own advantages/disadvantages (see e.g.; \citealp{Muldrew12,Darvish15a}). These measures include the halo mass, halo size, the local overdensity of galaxies, cluster or group membership, distance to the center of the parent halo, cluster, or group, association with different components of the cosmic web, and so on. Throughout this paper, we use the term ``environment'' or ``local environment'' to refer to the environment traced by the overdensity of galaxies.
   
\subsubsection{Local Universe} \label{method-env-sdss}
 
We use the projected comoving distance to the $10th$ nearest neighbor to each galaxy, considering only galaxies that are within the recessional velocity range of $\Delta v$=$c\Delta z$=$\pm$1000 kms$^{-1}$ to that galaxy, and corrected for incompleteness due to the fiber collision and flux limit of the sample:
\begin{equation} 
\Sigma_{i}= \frac{1}{C_{i}\Psi(z_{i})} \frac{10}{\pi d_{i}^{2}}
\end{equation}
where $\Sigma_{i}$ is the local projected surface density for the galaxy $i$, $d_{i}$ is the projected comoving distance to the $10th$ neighbor, $C_{i}$ is a correction term for the galaxy $i$ due to the spectroscopic fiber collision, and $\Psi(z_{i})$ is the selection function used to correct the sample for the Malmquist bias. 

$C_{i}$ is evaluated using the \cite{Baldry06} approach and is given in Appendix \ref{A} (see Figure \ref{fig:comp-den}). To estimate $\Psi(z_{i})$, we follow \cite{Efstathiou01} by modelling the change in the number of galaxies (in redshift bins of $\Delta z$=0.005) as a function of redshift with:
\begin{equation} 
N(z)dz=Az^{2}\Psi(z)dz, \quad \textrm{where} \quad \Psi(z)= e^{-(z/z_{c})^{\alpha}}
\end{equation}
where $A$ is a normalization factor, and $z_{c}$ is a characteristic redshift that corresponds to the peak of the redshift distribution. The best fitted model is given by $A$=8.50$\pm$0.75 $\times$ 10$^{6}$, $z_{c}$=0.0653$\pm$0.0035, and $\alpha$=1.417$\pm$0.054 (Figure \ref{fig:sf} (a) Appendix \ref{A}). To avoid large uncertainties and fluctuations in the estimated densities due to smaller sample size at higher redshifts, we only use galaxies for which $\Psi(z)$ $\geqslant$ 0.1 (Figure \ref{fig:sf} (b) Appendix \ref{A}). This corresponds to $z$ $\sim$ 0.12. For details of the method, why we use the distance to the $10th$ neighbor and the selection of $\Delta v$=$\pm$1000 kms$^{-1}$, see Appendix \ref{A}.

\subsubsection{High Redshift}
    
We use the density field estimation of \cite{Darvish17} in the COSMOS field. The local environment measurement relies on the adaptive kernel smoothing method \citep{Scoville13,Darvish15a} using a global kernel width of 0.5 Mpc, estimated over a series of overlapping redshift slices \citep{Darvish15a}. A mass-complete sample (similar to a volume-limited sample) is used for density estimation. There are several known large-scale structures (LSS) in the COSMOS field in the redshift range of our sample (e.g.; \citealp{Guzzo07,Finoguenov07,Sobral11,Scoville13,Darvish14}) which provide us with a relatively large dynamical range of environments for our high-$z$ sample at 0.6 $\lesssim$ $z$ $\lesssim$ 1.  

Using different density estimators at low- and high-$z$ ($10th$ nearest neighbor versus adaptive kernel smoothing) might lead to a potential bias in comparing the results at low and high redshift. However, in Appendix \ref{B}, we compare the density estimation using the $10th$ nearest neighbor and adaptive kernel smoothing for our high-$z$ sample and find a good agreement. Moreover, \cite{Darvish15a} find an overall good agreement between the estimated density fields using different methods (including the $10th$ nearest neighbor and adaptive kernel smoothing) over $\sim$ 2 dex in overdensity values through simulations and also observational data. Hence, the selection of different estimators has no significant effect on the presented results.
      
\subsection{Central and Satellite Selection} 

\subsubsection{Local Universe}

We rely on a sample of galaxy groups (in sample A) to select central and satellite galaxies. We select the brightest galaxy in each group as the central and the rest of group members as satellites. Galaxies that are not related to any galaxy group (isolated galaxies) are either centrals whose satellites, in principle, are too faint to be detected in our sample or they are ejected satellites moving beyond their halo's virial radius (e.g.; \citealp{Wetzel14}). Galaxy groups are selected using the friends-of-friends algorithm \citep{Huchra82}. Two galaxies $i$ and $j$ with redshifts $z_{i}$ and $z_{j}$ respectively and angular separation $\theta_{ij}$ are linked to each other if their projected ($D_{\perp,ij}$) and line-of-sight separations ($D_{\parallel,ij}$) satisfy the following conditions:
\begin{equation}
\begin{aligned}
D_{\perp,ij} &\leqslant b_{\perp} n(z)^{-1/3}, \quad  D_{\perp,ij} = \frac{c}{H_{0}} (z_{i}+z_{j})\sin(\theta_{ij}/2) \\
D_{\parallel,ij} &\leqslant b_{\parallel} n(z)^{-1/3}, \quad D_{\parallel,ij} = \frac{c}{H_{0}} |z_{i}-z_{j}| 
\end{aligned}
\end{equation}
where $c$ is the speed of light, $H_{0}$ is the Hubble constant, $n(z)$ is the mean number density of galaxies at $z$ (average redshift of galaxies $i$ and $j$) estimated from equation 2, and $b_{\perp}$ and $b_{\parallel}$ are the projected and line-of-sight linking lengths in units of the mean intergalaxy separation. Here, we use $b_{\perp}$=0.07 and $b_{\parallel}$=1.1 proposed by \cite{Duarte14} to be best suited for environmental studies. In Section \ref{science}, when we use the term ``all galaxies'', we mean all galaxies in our sample (central+satellite+isolated).

\subsubsection{High Redshift}   

We match our high-$z$ sample with \cite{Darvish17} catalog of satellites, centrals, and isolated systems in the COSMOS field. Their group selection is similar to that of our local universe sample but the linking parameters are optimized according to their selection functions. Nonetheless, the fraction of different galaxy types is very similar between the SDSS and COSMOS galaxies which guarantees a reliable comparison between our low- and high-$z$ samples (15(16)\%, 46(48)\%, and 39(36)\% for SDSS(COSMOS) centrals, satellites, and isolated systems, respectively). 

\subsection{Galaxy Physical Properties} \label{property}

\subsubsection{Method}

Our extraction of galaxy physical properties relies on the \cite{Martin17} method. It utilizes semi-analytical models \citep{Delucia06} in the context of the cosmological N-body simulation \citep{Springel05} to generate a sample of model galaxies at 0 $<$ $z$ $<$ 6 with known physical parameters such as, star-formation rate (SFR), stellar mass, and other parameters including the instantaneous time derivative of the star formation rate that we denote as the Star Formation Acceleration (SFA) and a similar quantity we denote as the Star Formation Jerk (SFJ). Single stellar populations \citep{Bruzual03} and a simple extinction slab model are then used to convert the star-formation histories into observable colors and spectral indices. At each D$_{n}$(4000) bin and redshift, a linear regression fit is then performed between the physical parameters and the model observables, resulting in a series of coefficients that are later used to convert the actual observables to the physical parameters for galaxy samples. The observables that we use here are the rest-frame $FUV-NUV$, $NUV-u$, $u-g$, $g-r$, $r-i$ colors, rest-frame M$_{i}$ absolute magnitude, D$_{n}$(4000), and H$\delta$:
\begin{eqnarray}
\notag
P_p(est) & =  & C_{1,p,d,z} (FUV-NUV) +  
\\  
\notag
& & C_{2,p,d,z} (NUV-u) + C_{3,p,d,z} (u-g) +  
\\
\notag
& &  C_{4,p,d,z} (g-r) + C_{5,p,d,z} (r-i) +
\\
\notag
& &  C_{6,p,d,z} H_{\delta} + C_{7,p,d,z} D_n(4000) + 
\\
& & C_{8,p,d,z} M_i + CTE_{p,d,z} 
\end{eqnarray}
where $P$ is the estimated physical parameter, $C_{i,p,d,z}$ is the coefficient of the observable $i$ for the physical parameter $p$ at redshift bin of $z$ and D$_{n}$(4000) bin of $d$, and $CTE_{p,d,z}$ is a constant. If the sources are not detected in the $FUV$ band, we only rely on other observables in determining the physical parameters \footnote{This is particularly important since quiescent galaxies and dusty systems may not have a high level of $FUV$ emission to be detected. Hence, exclusion of non-detected $FUV$ sources would automatically bias the analysis to samples with higher sSFR and low dust content.}.

The derived physical parameters that we use in this work are SFA (in units of mag Gyr$^{-1}$; defined as $\frac{d(NUV-i)_{0}}{dt}$ where $dt$ is the past 300 Myr and $(NUV-i)_{0}$ is the extinction-corrected $NUV-i$ color), SFJ (in units of mag Gyr$^{-1}$; defined as $\frac{d(NUV-i)_{0}}{dt}$ where $dt$ is the past 600-300 Myr), stellar mass $M_{*}$, and sSFR. A positive (negative) value of SFA and SFJ indicates quenching (bursting) in the past 300 Myr and the past 600-300 Myr, respectively. The combination of SFA and SFJ can place constraints on the strength and the typical timescale of quenching and bursting. 

\cite{Martin17} compared the derived $M_{*}$, SFRs, and other physical quantities in the local universe with similar ones in the literature and found a relatively good agreement (within $\sim$ 0.1-0.2 dex).
For details of the method, potential degeneracies, and comparisons with the literature, see \cite{Martin17}. Some other comparisons can be found in Appendix \ref{C} of this paper.
\begin{figure}
\centering
\includegraphics[width=3.5in]{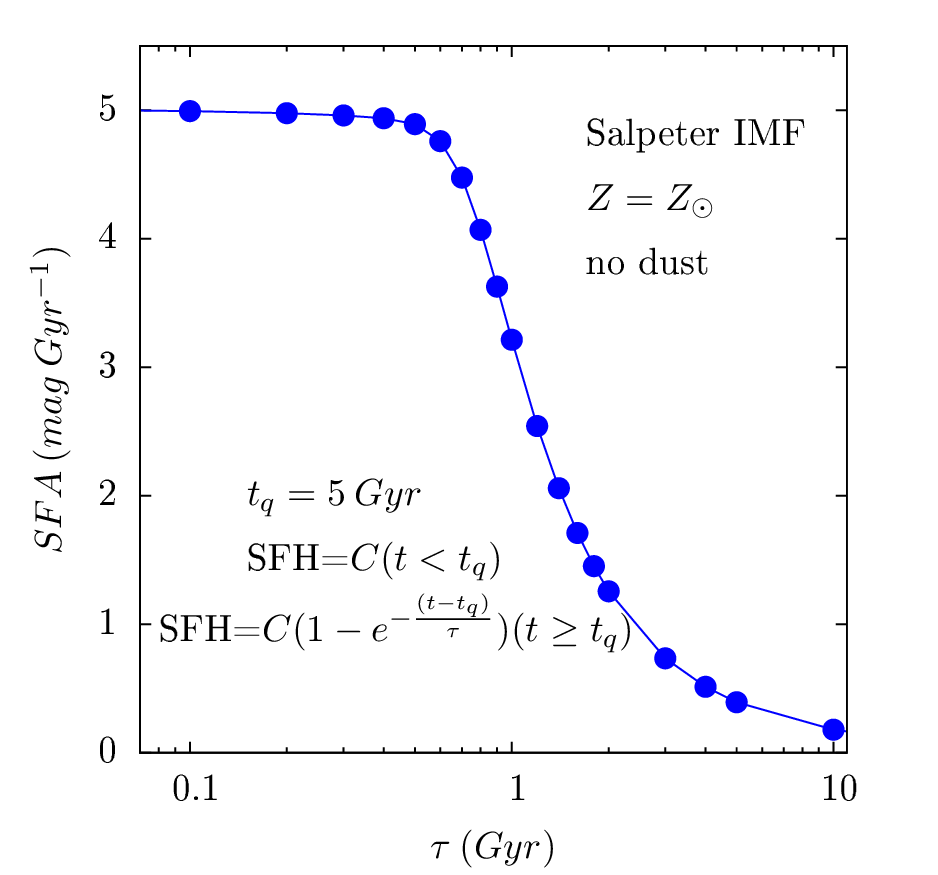}
\caption{SFA as a function of quenching timescale $\tau$ for a constant, 5 Gyr-long SFH, followed by an exponentially declining SFH with different quenching timescales $\tau$. Unextinguished models with Salpeter IMF and solar metallicity are used.}
\label{fig:timescale}
\end{figure}
\subsubsection{SFA and Quenching/Bursting timescale} 

In \cite{Martin17}, no prior assumptions are made about the shape of the star-formation histories (SFH) used in extracting the physical parameters. This allows us to extract new physical parameters such as SFA. However, in order to give a sense of how the SFA is related to the typical quenching/bursting timescales, we model the changes in $NUV-i$ color with time (used in the SFA definition) assuming an exponentially declining SFH with different e-folding (quenching) timescales \citep{Martin07}. We assume that the SFR is constant for 5 Gyr, followed by an exponentially declining SFH ($\propto$ e$^-\frac{t}{\tau}$) with different $\tau$ values. We model the $NUV-i$ color changes (SFA) after the onset of quenching using \cite{Bruzual03} models, assuming a Salpeter IMF, solar metallicity, and no dust. Figure \ref{fig:timescale} shows the SFA as a function of quenching timescale $\tau$ for this simplistic model. Note that the relation between SFA and $\tau$ should be used with caution given the assumptions used here. However, Figure \ref{fig:timescale} gives us a qualitative impression about the physical meaning of SFA that will be extensively used in the following section.   

\begin{figure*}
\centering
\includegraphics[width=7.0in]{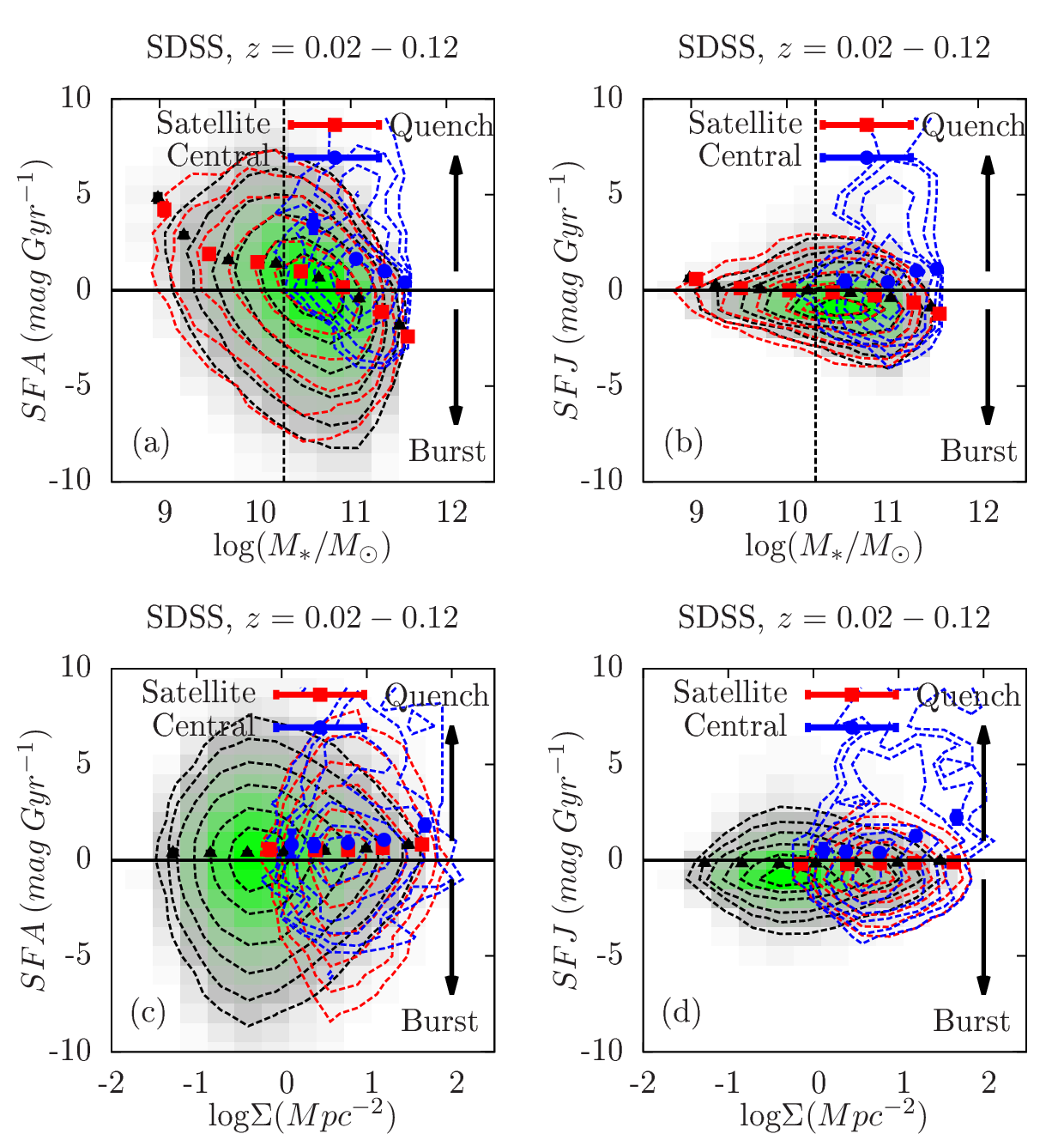}
\caption{(a) Median SFA as a function of stellar mass for all (black triangles), satellite (red squares), and central (blue circles) galaxies in the local universe. The overall distribution of SFA vs. $M_{*}$ is shown as a heat map. Black, red, and blue contours correspond to all, satellite, and central galaxies, respectively. Contour levels are at $3/4th$, $1/2th$, $1/4th$, $1/8th$, $1/16th$, and $1/32th$ of the peak. Black vertical line shows the stellar mass completeness limit. A positive (negative) SFA value indicates recent quenching (bursting) in the past 300 Myr. On average, less massive galaxies tend to be quenching and more massive systems bursting with a transition at log($M_{*}/M_{\odot}$) $\sim$ 10.5-11. Satellites follow the general trends between SFA and $M_{*}$. Centrals avoid the bursting region and at a given $M_{*}$, centrals are quenchier than satellites. (b) Similar to (a) but for SFJ vs. $M_{*}$. A positive (negative) value indicates quenching (bursting) at 600-300 Myr prior to observations. A very weak correlation between SFJ and $M_{*}$ is seen. (c) Median SFA as a function of local density for all, satellite, and central galaxies in the local universe, with no (or a weak) environmental dependence when averaged over all stellar masses. (d) Similar to (c) but for SFJ vs. log$\Sigma$.}
\label{fig:mass-density-SFA-SFJ}
\end{figure*}

\begin{figure*}
\centering
\includegraphics[width=7.0in]{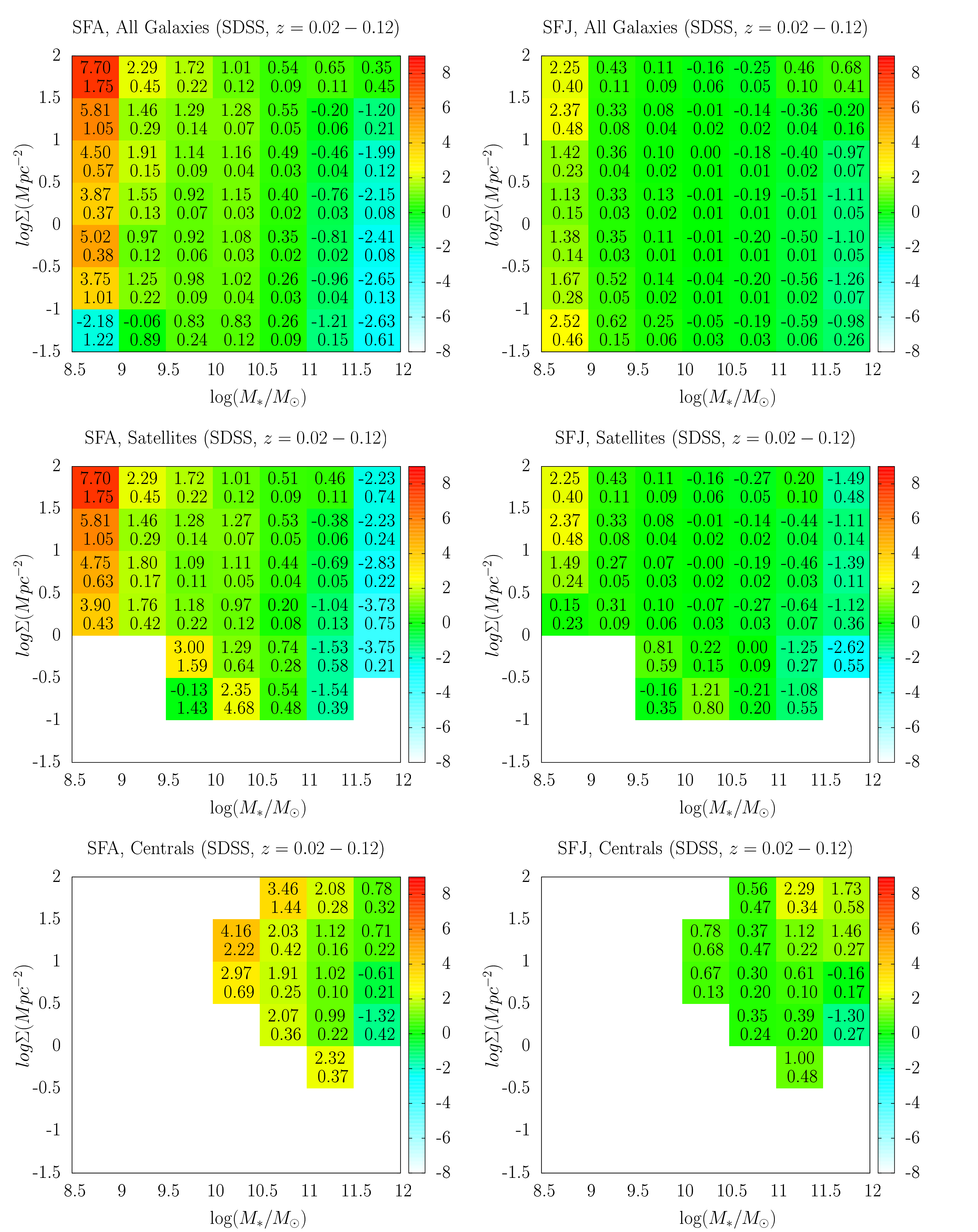}
\caption{Median SFA (left) and SFJ (right) shown by color on the diagram of log$\Sigma$ versus log($M_{*}/M_{\odot}$) for all galaxies (top), satellites (middle), and centrals (bottom) in our SDSS sample at $z$ $\sim$ 0. The top number in each cell is the median value (SFA or SFJ) and the bottom one is its uncertainty. In each environment, more massive systems are burstier than less massive ones. The local environmental dependence of SFA is also evident,i.e.; in each mass bin and on average, denser environments host higher quenchiness than the less-dense field. Note that although the SFA depends on both $M_{*}$ and $\Sigma$, the stellar mass dependence is stronger. Also note that the environmental dependence of SFA is less significant in the medium range of stellar masses (log($M_{*}/M_{\odot}$) $\approx$ 9.5-11). The SFA of satellites also depends on both $M_{*}$ and $\Sigma$. However, the SFA of centrals only shows a mass dependence and within the uncertainties, it is almost independent of the local environment (or at best has a weak dependence). Note that in each stellar mass and local environment bin, centrals are quenchier than satellites in an average sense, and that centrals are mainly quenching. Compared to the SFA, the SFJ shows much weaker dependence on stellar mass and almost no (or at best a weak) environmental dependence.}
\label{fig:2D-MD}
\end{figure*}

\begin{figure*}
\centering
\includegraphics[width=7.0in]{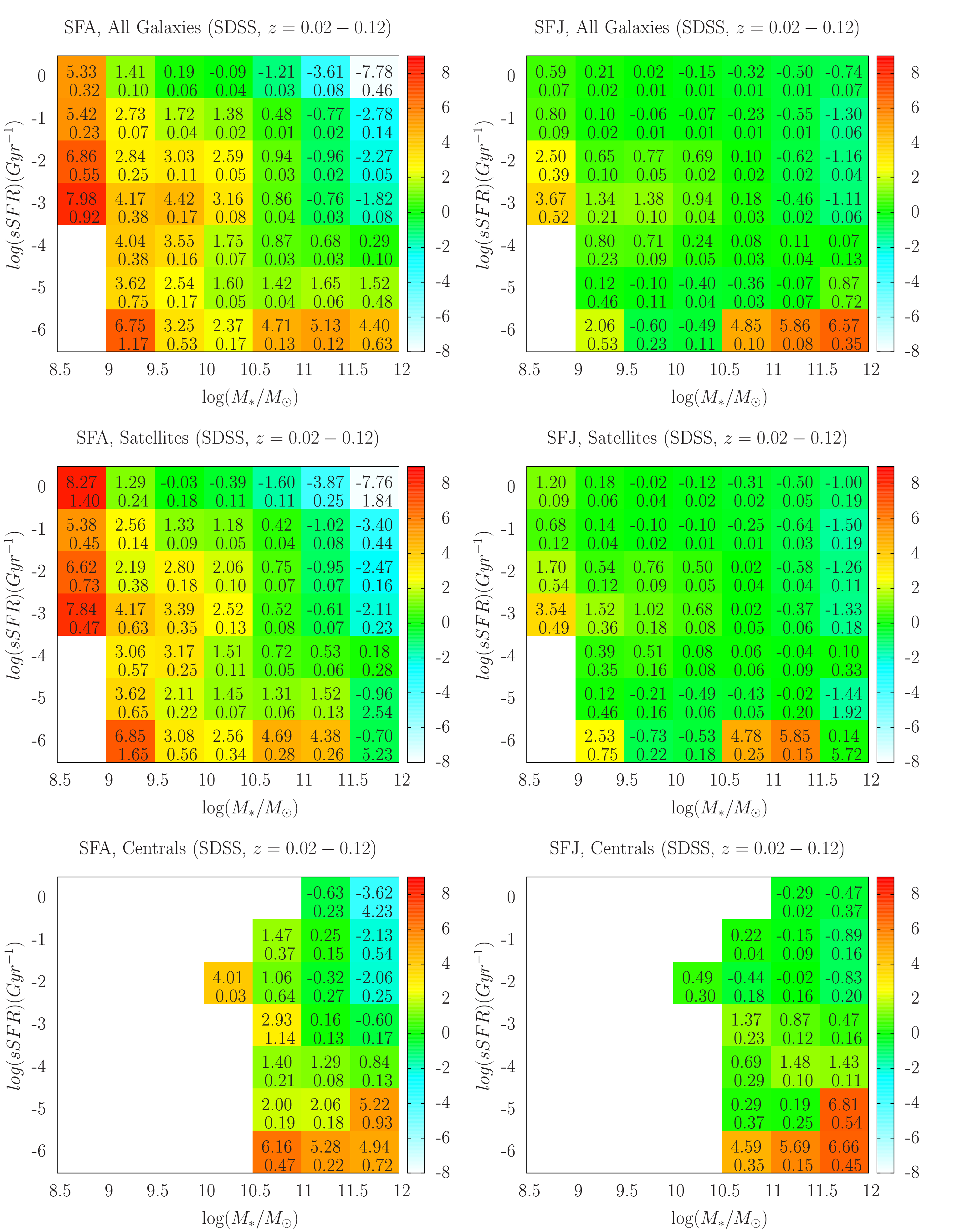}
\caption{Median SFA (left) and SFJ (right) shown by color on the diagram of log(sSFR) versus log($M_{*}/M_{\odot}$) for all galaxies (top), satellites (middle), and centrals (bottom) in our SDSS sample at $z$ $\sim$ 0. The top number in each cell is the median value (SFA or SFJ) and the bottom one is its uncertainty. At fixed sSFR and on average, less massive galaxies are quenchier than more massive systems. The SFA strongly depends on sSFR as well. At fixed stellar mass and on average, the median SFA increases with decreasing sSFR. The burstiness happens in massive star-forming galaxies (log($M_{*}/M_{\odot}$) $\gtrsim$ 11) with high sSFR values (log(sSFR)(Gyr$^{-1}$) $\gtrsim$ -3). On average, the SFA decreases with increasing $M_{*}$ and sSFR for satellites and centrals as well. However, in each $M_{*}$ and sSFR bin, centrals are quenchier than (or have similar SFA to) satellites. Compared to SFA, the SFJ shows weaker dependence on $M_{*}$ and sSFR (or at best similar values in massive, low-sSFR galaxies).}
\label{fig:2D-MsSFR}
\end{figure*}

\begin{figure*}
\centering
\includegraphics[width=7.0in]{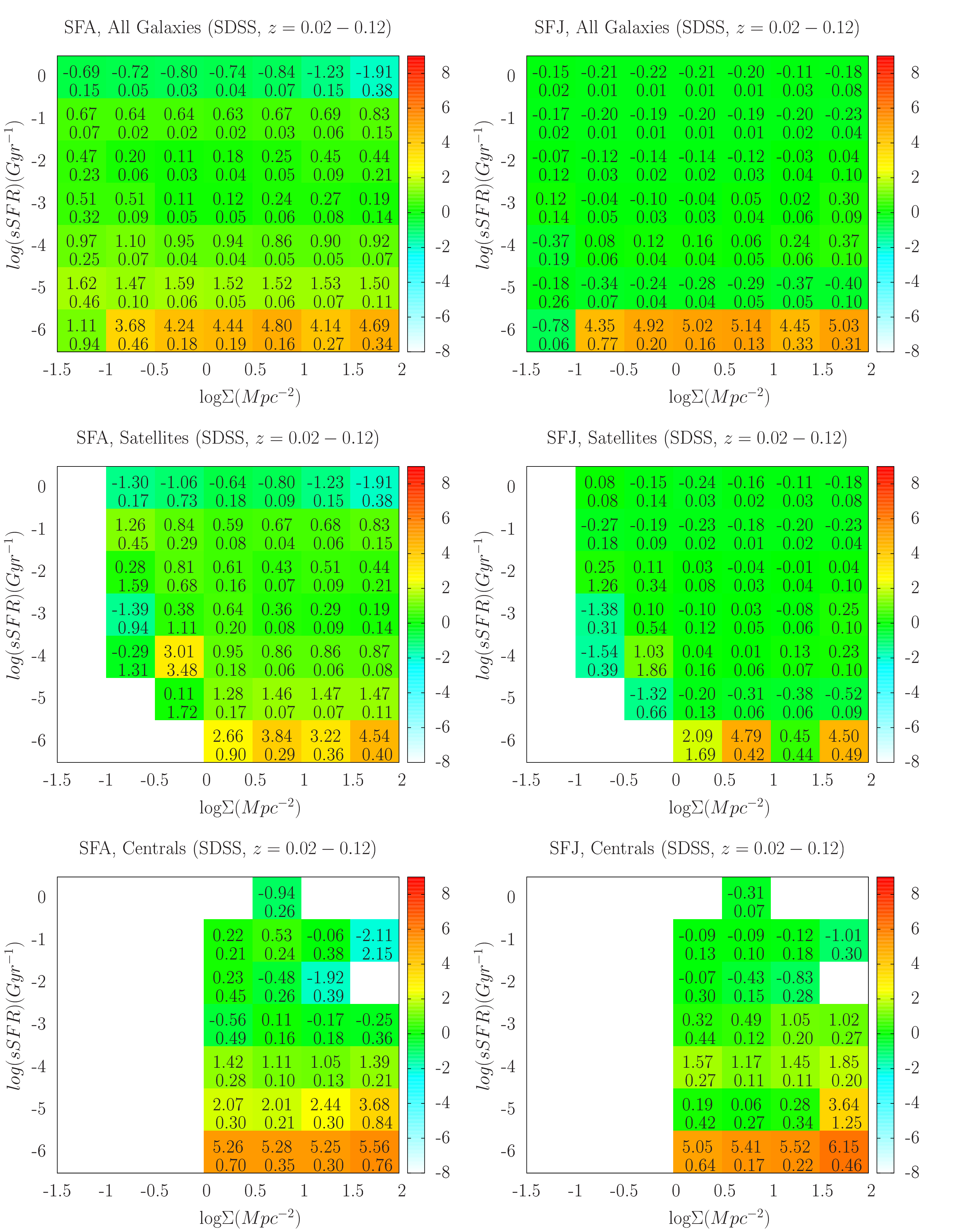}
\caption{Median SFA (left) and SFJ (right) shown by color on the plane of log(sSFR) versus log$\Sigma$ for all galaxies (top), satellites (middle), and centrals (bottom) in our SDSS sample at $z$ $\sim$ 0. The top number in each cell is the median value (SFA or SFJ) and the bottom one is its uncertainty. At fixed environment, the median SFA depends on sSFR and increases with decreasing sSFR. However, at fixed sSFR and within the uncertainties, the median SFA is almost independent of $\Sigma$. These results hold for all galaxies, satellites, and centrals. The weaker sSFR and $\Sigma$ dependence of the SFJ compared to SFA is also seen.}
\label{fig:2D-DsSFR}
\end{figure*}

\begin{figure*}
\centering
\includegraphics[width=7.0in]{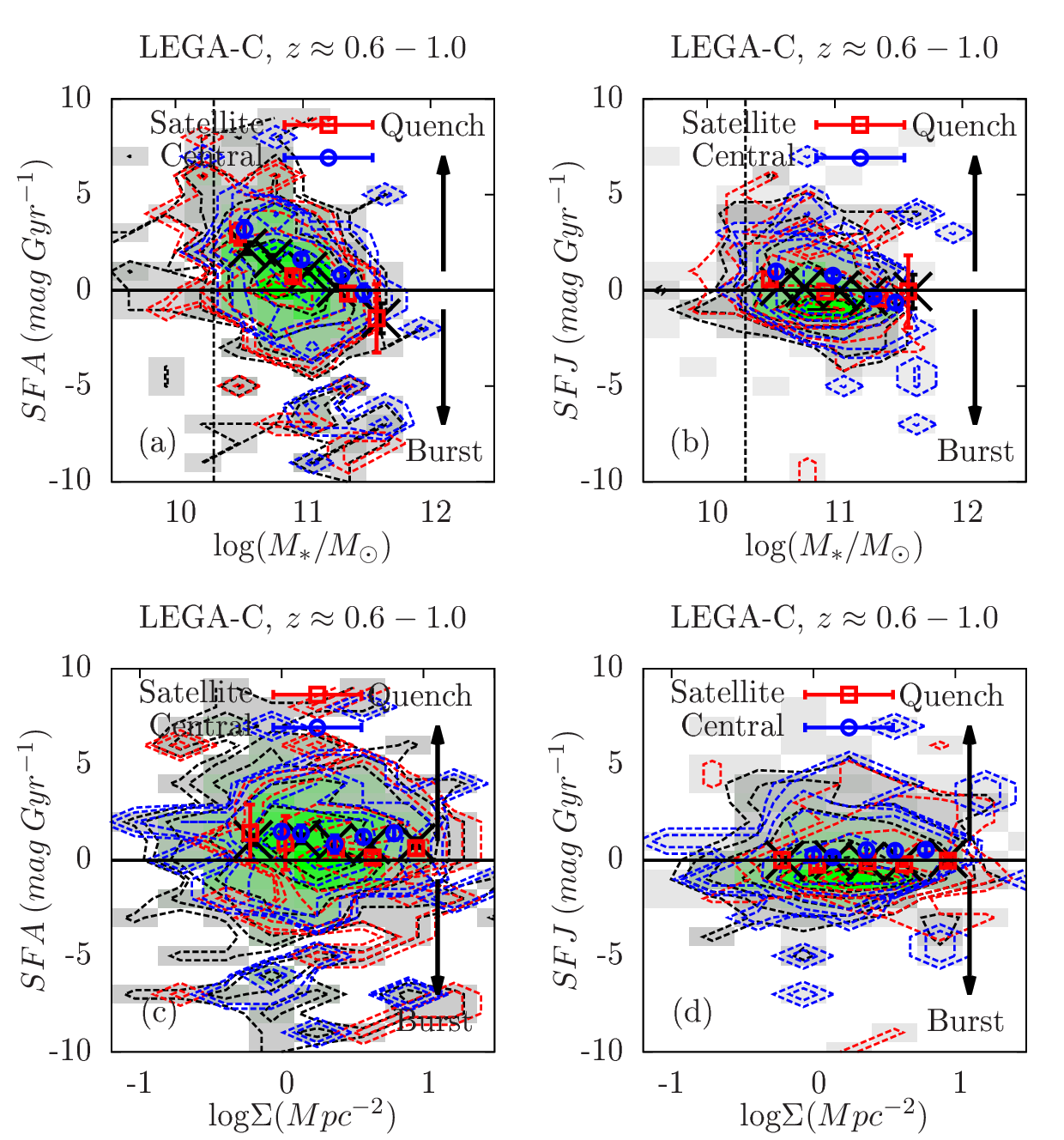}
\caption{Similar to Figure \ref{fig:mass-density-SFA-SFJ} but for our LEGA-C high-$z$ sample at $z$ $\sim$ 1. Red, blue, and black points show the median values for satellite, central, and all galaxies. Similar to SDSS results, SFA (and SFJ to a lesser degree) decreases with increasing $M_{*}$ for all galaxies, satellites, and centrals, with evidence for centrals being quenchier than satellites at fixed $M_{*}$. We also find an environmental independence (when averaged over all the masses and out to log$\Sigma$ $\sim$ 1) of the SFA and SFJ at $z$ $\sim$ 1, similar to our results in the local universe.}
\label{fig:hz-mass-density-SFA-SFJ}
\end{figure*}

\begin{figure*}
\centering
\includegraphics[width=7.0in]{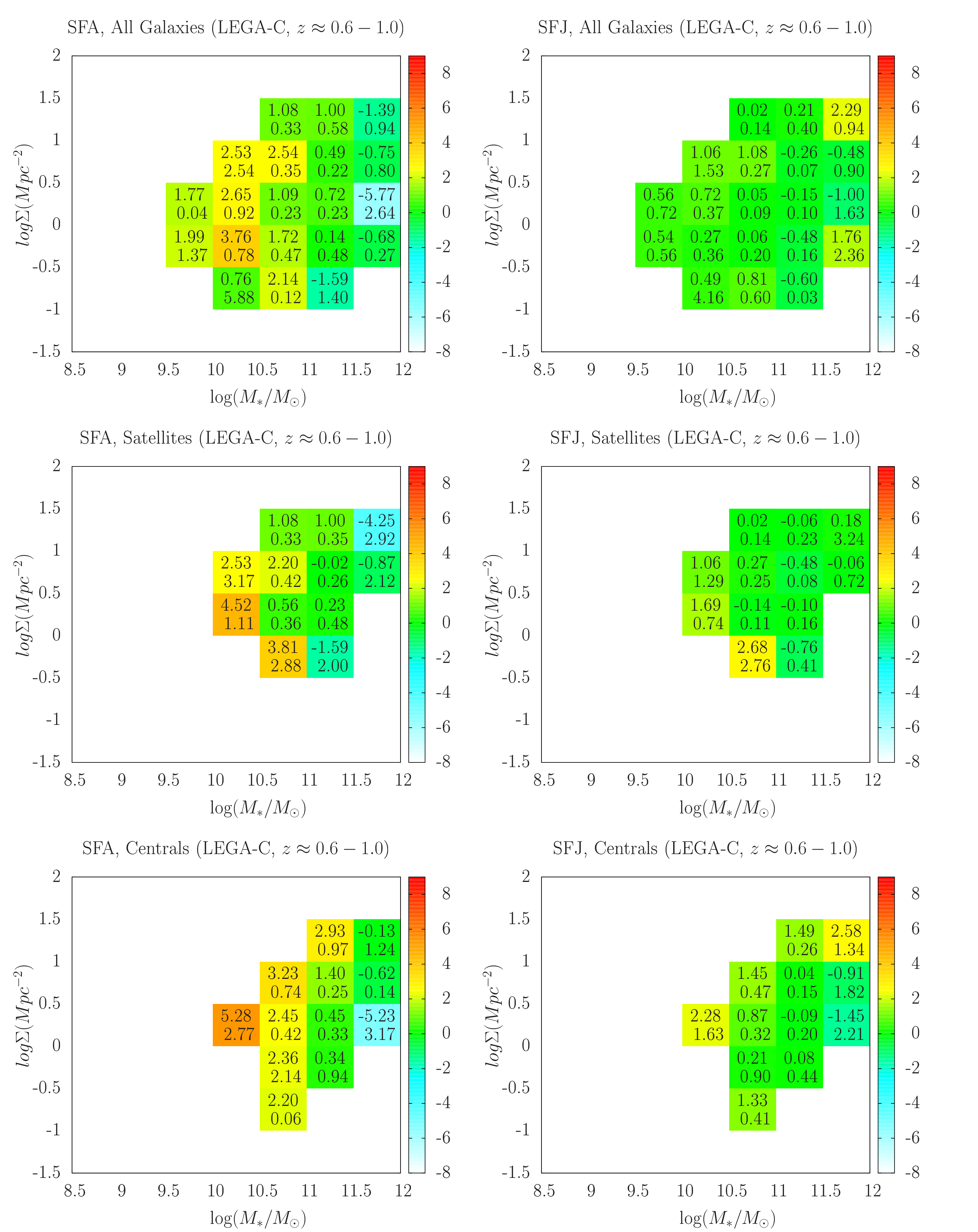}
\caption{Similar to Figure \ref{fig:2D-MD} but for our LEGA-C sample at $z$ $\sim$ 1. The median SFA decreases with increasing stellar mass (even at fixed environment) at $z$ $\sim$ 1. Unfortunately, due to a smaller dynamical range of environment, stellar mass, and sample size and larger uncertainties in our high-$z$ sample compared to the local universe, we cannot make a significantly robust statement about the potential SFA relation with environment (at a given stellar mass). The overall sample of galaxies shows signs of increasing SFA (and quenching) in denser environments at fixed $M_{*}$. However, because of a smaller dynamical range of $\Sigma$, $M_{*}$, and sample size and larger uncertainties than the low-$z$ sample, we cannot make a robust statement on the potential environmental dependence of SFA (at given stellar mass) at $z$ $\sim$ 1. The stellar mass dependence of SFA is also seen for both satellites and centrals at $z$ $\sim$ 1 but within the uncertainties and in the $M_{*}$ and $\Sigma$ range covered at high $z$, no clear relation between SFA and environment is seen for centrals and satellites. Even at $z$ $\sim$ 1, centrals are quenchier than satellites on average. Similar to the SDSS results, the SFJ shows weaker dependence on $M_{*}$ and $\Sigma$ (if any) than the SFA.} 
\label{fig:hz-2D-MD}
\end{figure*}

\begin{figure*}
\centering
\includegraphics[width=7.0in]{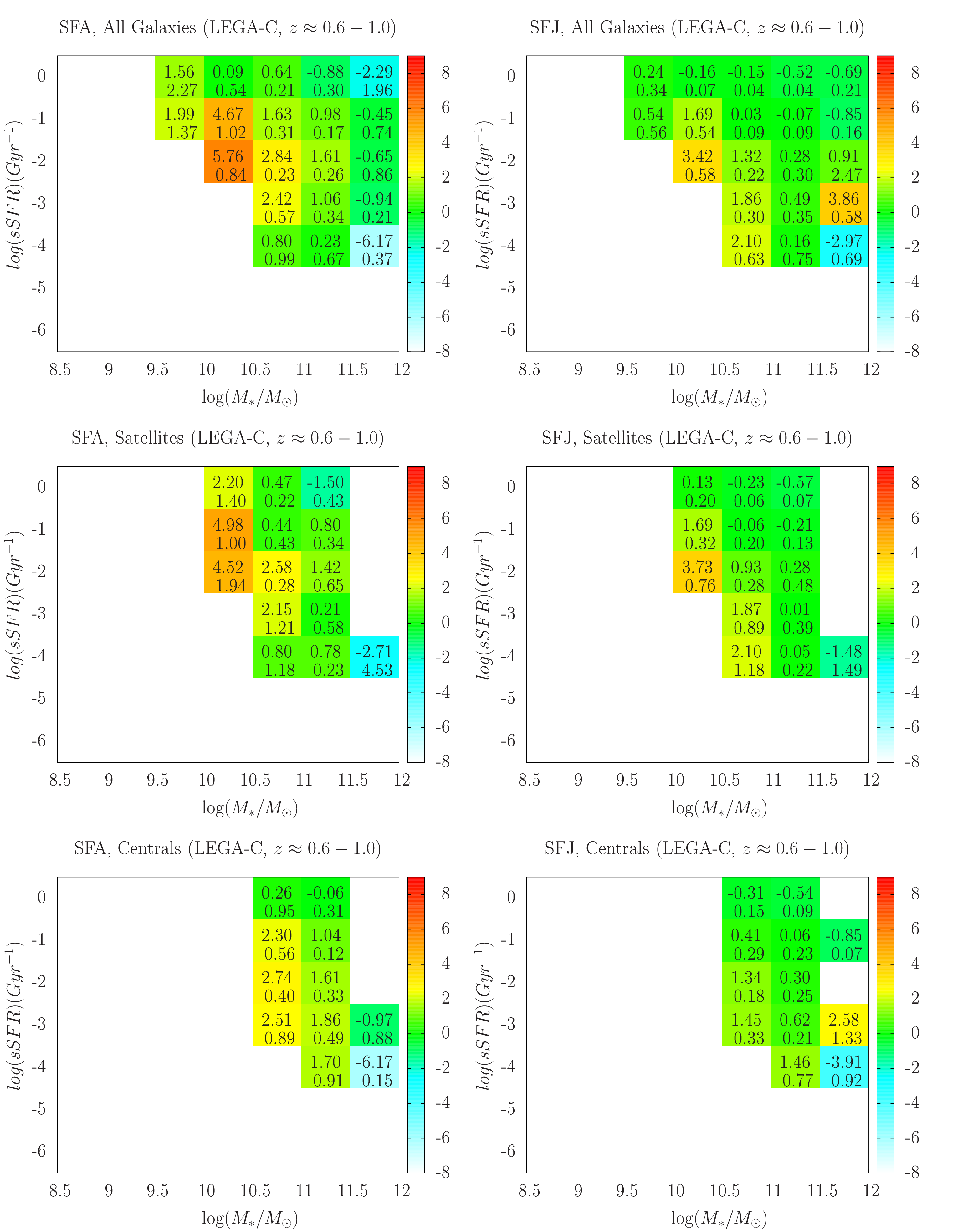}
\caption{Similar to Figure \ref{fig:2D-MsSFR} but for our high-$z$ LEGA-C sample at $z$ $\sim$ 1. Similar to the low redshift results, the median SFA decreases with increasing stellar mass and sSFR for all galaxies, satellites and centrals. Similarly, the SFJ shows weaker (or similar) dependence on stellar mass and sSFR than the SFA.}
\label{fig:hz-2D-MsSFR}
\end{figure*}

\begin{figure*}
\centering
\includegraphics[width=7.0in]{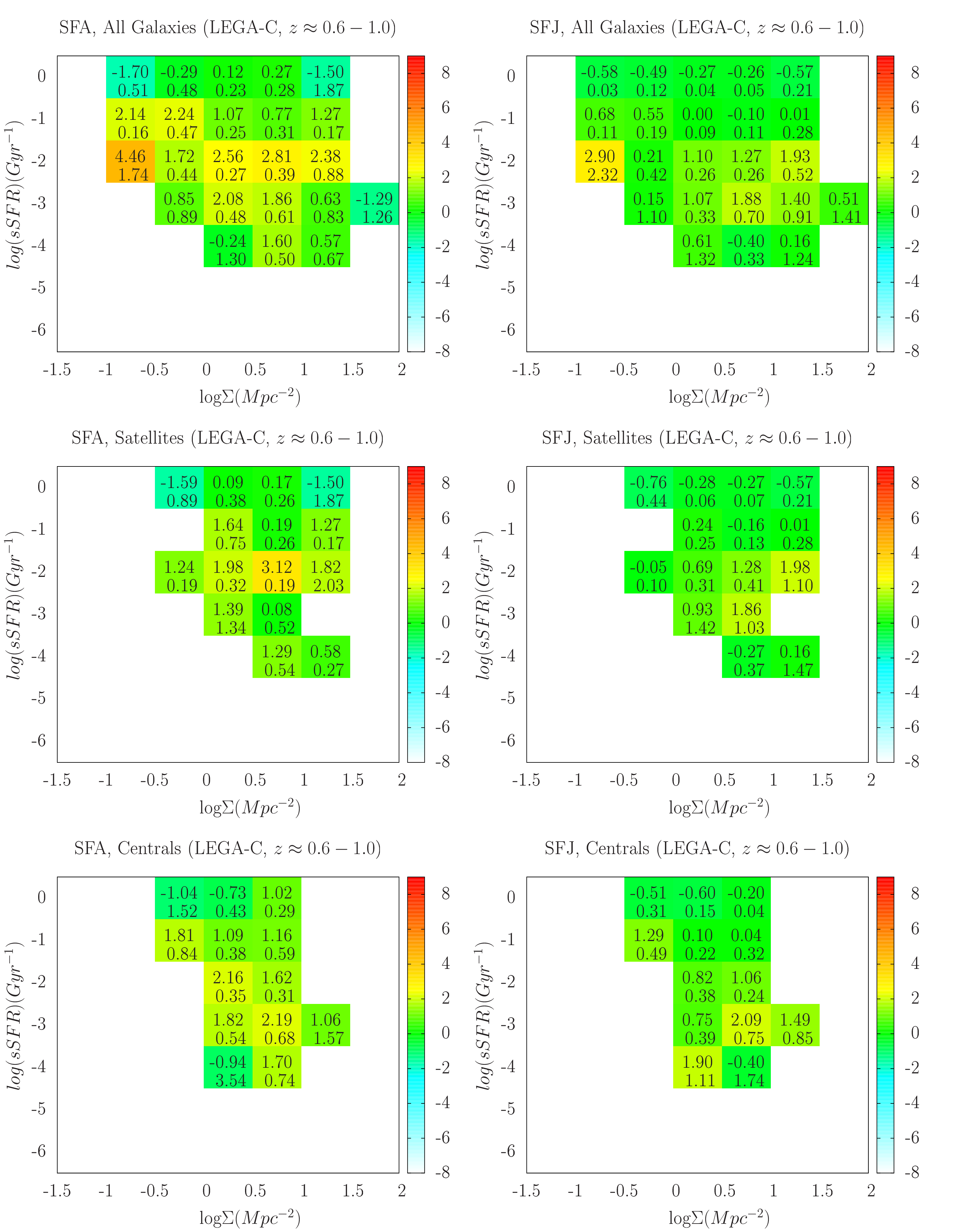}
\caption{Similar to Figure \ref{fig:2D-DsSFR} but for our sample at $z$ $\sim$ 1. Given the small range of environments and sSFRs, and large uncertainties, no significant trend between SFA and environment or sSFR is seen here. However, some environmental bins show signs of an increasing SFA with decreasing sSFR, similar to the SDSS results. At fixed sSFR and $\Sigma$, the SFJ shows smaller values than (or similar values to) the SFA.}
\label{fig:hz-2D-DsSFR}
\end{figure*}

\begin{figure}
\centering
\includegraphics[width=3.5in]{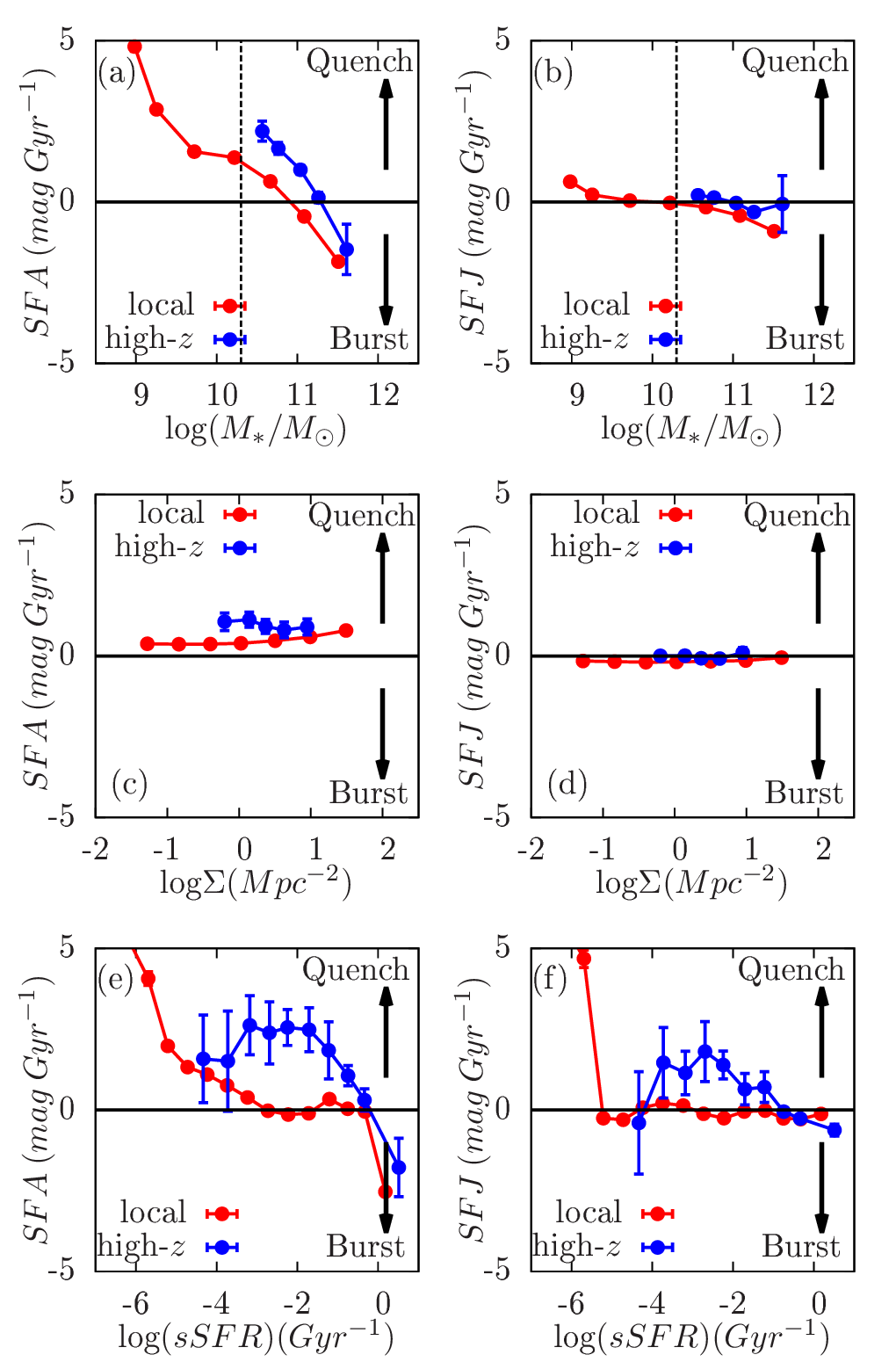}
\caption{(a) Median SFA vs. $M_{*}$ (b) SFJ vs. $M_{*}$ (c) SFA vs. log$\Sigma$ (d) SFJ vs. log$\Sigma$ (e) SFA vs. sSFR and (f) SFJ vs. sSFR for our local universe and high-$z$ samples. On average, at fixed $M_{*}$, sSFR, and (to a lesser degree) $\Sigma$, the higher redshift galaxies are quenchier. Black vertical line shows the stellar mass completeness limit.}
\label{fig:evo-mass-density-SFA-SFJ}
\end{figure}
 
\section{Results} \label{science}

\subsection{Quenching/Bursting of Galaxies in the Local Universe} \label{science-lowz}

Figure \ref{fig:mass-density-SFA-SFJ} (a) shows the SFA as a function of stellar mass for our SDSS sample (black triangles). A positive (negative) value indicates recent quenching (bursting) in the past 300 Myr. We clearly see a trend with stellar mass, in the sense that on average, less massive galaxies tend to be quenching and more massive systems bursting, consistent with \cite{Martin17}. The transition between quenching and bursting occurs at log($M_{*}/M_{\odot}$) $\sim$ 10.5-11. Figure \ref{fig:mass-density-SFA-SFJ} (b) shows the SFJ versus stellar mass for our local universe sample. A positive (negative) value indicates past quenching (bursting) at 600-300 Myr prior to observations. We still see a very weak correlation between SFJ and $M_{*}$ particularly at log($M_{*}/M_{\odot}$) $\gtrsim$ 11 but clearly, much of the quenching/bursting has happened recently as seen in Figure \ref{fig:mass-density-SFA-SFJ} (a). This indicates that the physics of mass quenching/bursting acts in a relatively short timescale ($\lesssim$ 300 Myr).

Figures \ref{fig:mass-density-SFA-SFJ} (a) and (b) also show the SFA and SFJ versus $M_{*}$ for central and satellite galaxies. To minimize the projection and group selection effects and contamination by interlopers, we only consider satellites and centrals that are in groups with $>$ 10 members. Satellites follow the general trends between SFA and SFJ versus $M_{*}$. Centrals follow the same slope between SFA and $M_{*}$. \textbf{However, centrals tend to avoid the bursting region and at a given $M_{*}$, centrals are quenchier than satellites.} 

Figures \ref{fig:mass-density-SFA-SFJ} (c) and (d) show the role of the local environment ($\Sigma$) on the SFA and SFJ. When averaged over all stellar masses, we find no clear trend (at best a weak correlation) between SFA (or SFJ) and $\Sigma$. Except for an increasing SFJ for centrals in dense regions, satellites and centrals do not show any significant environmental dependence in their very recent ($<$ 300 Myr) and less recent (past 300-600 Myr) quenching/bursting as denoted by SFA and SFJ quantities (when averaged over all stellar masses). This indicates that local environment likely acts effectively on a much longer timescale when averaged over all $M_{*}$. There are other possibilities too. For example, the local environment might not affect the quantities that are linked to quenching/bursting of galaxies. It might also be due to the mass quenching/bursting being more effective than the environmental quenching/bursting when averaged over the general population of galaxies.

We further investigate the quenching/bursting of galaxies by dividing our sample into stellar mass, sSFR, and density bins. Figure \ref{fig:2D-MD} shows the median SFA and SFJ (shown by color) on the log$\Sigma$ vs. log($M_{*}/M_{\odot}$) plane for all galaxies, satellites, and centrals. The top number in each cell is the median value and the bottom one is its uncertainty. The mass dependence of SFA is clearly seen on the log$\Sigma$ versus log($M_{*}/M_{\odot}$) diagram, in a sense that in any given environment, more massive systems are burstier than less massive galaxies. However, the local environmental dependence of SFA is also evident. In each mass bin, on average, denser environments host higher quenchiness than the less-dense field. The largest burstiness occurs in massive field galaxies (log($M_{*}/M_{\odot}$) $\gtrsim$ 11.5 and log$\Sigma$ $\lesssim$ 0) and the largest quenchiness belongs to low-mass systems in very dense environments (log($M_{*}/M_{\odot}$) $\lesssim$ 9.0 and log$\Sigma$ $\gtrsim$ 0.5). 

In addition, at fixed $\Sigma$, the SFA change with stellar mass is stronger than the SFA change with environment while fixing $M_{*}$. In other words, although the SFA depends on both stellar mass and environment, the stellar mass dependence is stronger. Also note that the environmental dependence of SFA is less significant in the medium range of stellar masses (log($M_{*}/M_{\odot}$) $\approx$ 9.5-11) and that is why on average, we do not find a significant environmental dependence of SFA in Figure \ref{fig:mass-density-SFA-SFJ}. 

Median SFA of satellites follows the general distribution of galaxies and it depends on both stellar mass and environment. However, SFA of centrals only shows a mass dependence and within the uncertainties, it is almost independent of the local environment (or at best has a weak dependence). This suggests that the environmental dependence of SFA is mostly due to satellites. 

Note that in each stellar mass and environment bin, centrals are quenchier than satellites and that centrals are mainly quenching. Compared to the SFA, the SFJ shows much weaker dependence on $M_{*}$ and almost no (or at best a weak) environmental dependence.

The strong sSFR dependence of the trends investigated so far is also evident in Figures \ref{fig:2D-MsSFR} and \ref{fig:2D-DsSFR}. Figure \ref{fig:2D-MsSFR} shows the median SFA and SFJ on the diagram of log(sSFR) versus log($M_{*}/M_{\odot}$) for all galaxies, satellites, and centrals at $z$ $\sim$ 0. The mass dependence of SFA is also seen on the log(sSFR) versus log($M_{*}/M_{\odot}$) diagram, i.e.; at fixed sSFR and on average, more massive galaxies are burstier than less massive systems. However, the SFA strongly depends on sSFR as well. At fixed stellar mass and on average, the median SFA increases with decreasing sSFR, confirming the $M_{*}$-sSFR trend seen in \cite{Martin17} for ``all galaxies''. In fact, the burstiness occurs in massive (log($M_{*}/M_{\odot}$) star-forming galaxies with high sSFRs (log(sSFR)(Gyr$^{-1}$) $\gtrsim$ -3). 

Similar trends are seen for satellites and centrals, that is, on average, the SFA increase with decreasing stellar mass and sSFR for both satellites and centrals. However, in each $M_{*}$ and sSFR bin, centrals quench faster than (at best have similar SFA to) satellites. Compared to SFA, the SFJ shows weaker dependence on $M_{*}$ and sSFR (at best similar values in massive, low-sSFR galaxies).

Finally, the SFA and SFJ as a function of sSFR and local density is shown in Figure \ref{fig:2D-DsSFR}. At fixed environment, the median SFA depends on sSFR and increases with decreasing sSFR. However, at fixed sSFR and within the uncertainties, the median SFA is almost independent of the local density of galaxies. These results hold for all galaxies, as well as satellites and centrals. The weaker sSFR and $\Sigma$ dependence of the SFJ compared to SFA is also seen. \textbf{Combining the results in Figures \ref{fig:2D-MD}, \ref{fig:2D-MsSFR}, and \ref{fig:2D-DsSFR} indicates that at $z$ $\sim$ 0, much of the bursting of star-formation happens in massive, high sSFR galaxies, particularly those in the field (and among group galaxies, satellites more than centrals), whereas most of the quenching of star-formation happens in less-massive, low sSFR galaxies, in particular those located in dense environments. For centrals, quenching is significant even in higher mass systems.}

\subsection{Quenching/Bursting of Galaxies at High-$z$} \label{science-highz}
 
We find similar results at $z$ $\sim$ 1 for our LEGA-C sample as shown in Figures \ref{fig:hz-mass-density-SFA-SFJ}, \ref{fig:hz-2D-MD}, \ref{fig:hz-2D-MsSFR}, and \ref{fig:hz-2D-DsSFR}. To increase the statistics, all satellites and centrals (with number of group members $\geqslant$ 2) are included \footnote{We note that because of this selection, the high-$z$ group galaxies are more prone to contamination by interlopers. However, in Appendix \ref{D}, we show that the overall trends are still retrieved (with larger uncertainties) using groups with $>$ 10 members.}. Also note the smaller dynamical range of the environment and $M_{*}$ probed here compared to that of the SDSS. Even with these limitations, some trends between the SFA (and to a smaller degree, the SFJ) and $M_{*}$, sSFR, and to a lesser degree environment are evident (with evidence of deviation between centrals and satellites). For example, as shown in Figure \ref{fig:hz-mass-density-SFA-SFJ}, we find a mass dependence and an environmental independence (when averaged over all $M_{*}$s and to log$\Sigma$ $\sim$ 1) of the SFA and SFJ at $z$ $\sim$ 1, with centrals being quenchier than satellites on average.

According to Figure \ref{fig:hz-2D-MD}, the decrease in median SFA with stellar mass (even at fixed environment) is also seen at $z$ $\sim$ 1. Unfortunately, due to a smaller dynamical range of $\Sigma$, $M_{*}$, and sample size and larger uncertainties in our high-$z$ sample compared to the local universe, we cannot make a significantly robust statement about the potential SFA relation with environment (at given stellar mass). However, even with these limitations, the overall sample of galaxies shows signs of increasing SFA (and quenching) in denser environments at fixed $M_{*}$. The stellar mass dependence of SFA is also seen for both satellites and centrals at $z$ $\sim$ 1 but within the uncertainties and in the stellar mass and environment range covered at high-$z$, no clear relation between SFA and environment is seen when we further break the sample into centrals and satellites. Note that even at $z$ $\sim$ 1, centrals seem to be quenchier than satellites in an average sense. Similar to the low-$z$ results, the SFJ shows weaker $M_{*}$ and $\Sigma$ dependence (if any) than the SFA. More importantly, we find that at given bins of stellar mass and environment, higher $z$ galaxies are on average quenchier than (or within the uncertainties, have at best similar SFAs to) their local-universe counterparts.

Similar to Figure \ref{fig:2D-MsSFR}, the median SFA decreases with increasing stellar mass and (and to a lesser degree) sSFR for all galaxies, satellites and centrals at $z$ $\sim$ 1. Moreover, at any given stellar mass and sSFR, higher redshift galaxies (all, centrals, and satellites) are on average quenchier than (or within uncertainties, have similar SFAs to) their local-universe counterparts. Similar to low-$z$ results, the SFJ shows weaker (or similar) trends with $M_{*}$ and sSFR than the SFA. Given the narrow range of environments and sSFRs, and large uncertainties, no significantly clear trend between SFA and environment or sSFR is seen in Figure \ref{fig:hz-2D-DsSFR}. However, some environmental bins show signs of an increasing SFA with decreasing sSFR, similar to the results at low-$z$. Moreover, at any given $\Sigma$ and sSFR, higher $z$ galaxies (all, centrals, and satellites) are on average quenchier than (or within uncertainties, have similar SFAs to) their local-universe counterparts. At fixed sSFR and $\Sigma$, the SFJ shows smaller values than (or similar values to) the SFA. 

In selecting the group galaxies for the high-$z$ sample, all groups with $\geqslant$ 2 members are considered. This makes the sample more prone to contamination by interlopers and might lead to unwanted biases when we compare the low- and high-$z$ results. However, in Appendix \ref{D}, we show that the global trends could still be recovered using groups with $>$ 10 members for the high-$z$ sample.     

\begin{figure}
\centering
\includegraphics[width=3.5in]{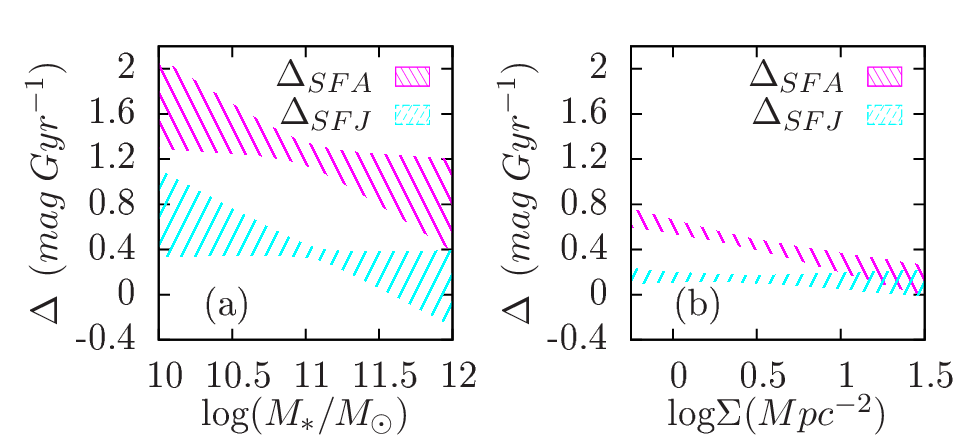}
\caption{$\Delta_{SFA}$=$\lvert SFA(z\sim 1)-SFA(z\sim 0)\rvert$ and $\Delta_{SFJ}$=$\lvert SFJ(z\sim 1)-SFJ(z\sim 0)\rvert$ as a function of stellar mass (a) and local environment (b). $\Delta$ is defined to investigate the redshift evolution of mass and environmental quenching/bursting. $\Delta_{SFA}$ (and to a lesser degree $\Delta_{SFJ}$) values show that the strength of the recent mass and environmental quenching/bursting is larger at higher redshift.}
\label{fig:evo-diff}
\end{figure}

\subsection{Redshift Evolution of Galaxy Quenching/Bursting}

\textbf{Comparing the low- and high-$z$ results indicates that at fixed $M_{*}$, sSFR, and environment, higher redshift galaxies (all, centrals, and satellites) are on average quenchier than their local-universe counterparts.} This is more clearly seen in Figure \ref{fig:evo-mass-density-SFA-SFJ}. At fixed $M_{*}$ (averaged over all environments and sSFRs, top panel), fixed sSFR (averaged over $\Sigma$ and $M_{*}$, bottom panel), and more slightly at fixed $\Sigma$ (averaged over sSFR and $M_{*}$, middle panel), on average, our high-$z$ sample is quenchier than the local-universe sample.

To further quantify the redshift evolution of the mass and environmental quenching and potentially compare their relative strength, we define the quantity $\Delta_{SFA}$=$\lvert SFA(z\sim 1)-SFA(z\sim 0)\rvert$ as the absolute difference between the median SFA at $z\sim$ 1 (LEGA-C sample) and the median SFA at $z\sim$ 0 (SDSS sample). We define a similar quantity for the SFJ ($\Delta_{SFJ}$=$\lvert SFJ(z\sim 1)-SFJ(z\sim 0)\rvert$). This absolute difference is done as a function of both stellar mass and environment and at the common stellar mass and environmental range of values for the high-$z$ and local-universe samples (log($M_{*}/M_{\odot}$) $\approx$ 10-12 and log($\Sigma$)$\approx$ -0.25-1.5). We model the median SFA(and SFJ) as a function of $M_{*}$ and $\Sigma$ with linear functions, taking their uncertainties into account. The absolute difference defined in $\Delta$ represents the difference between the linearly-modelled SFA (or SFJ). Figure \ref{fig:evo-diff} shows the results. Both stellar mass and environment show stronger strength in recent quenching/bursting of galaxies at higher redshift. Moreover, the change in stellar mass quenching/bursting seems to be larger ($\Delta_{SFA}(M_{*})$ $\approx$ 1.7-0.8 magGyr$^{-1}$ at 10 $\lesssim$log($M_{*}/M_{\odot}$)$\lesssim$ 12) than the environment ($\Delta_{SFA}(\Sigma)$ $\approx$ 0.4-0.1 magGyr$^{-1}$ at 0.5 $\lesssim$log($\Sigma$)$\lesssim$ 1.5) since $z$ $\sim$ 1. The SFJ shows weaker trends compared to the SFA.       

\section{Discussion} \label{dis}

To $z$ $\sim$ 1, we find that: 1- the SFA (and SFJ to a lesser degree) decreases with increasing stellar mass, increasing sSFR, and decreasing local density. 2- on average, centrals quench faster than satellites, and 3- high redshift galaxies are quenchier than their local-universe counterparts.
     
To explain the mass dependence of SFA, \cite{Martin17} proposed a scenario in which lower mass galaxies accrete into halos and become satellites, having their star-forming gas tidally and/or ram-pressure stripped, while higher mass centrals receive the gas and react with new star formation. However, by explicitly breaking the sample into satellites and centrals, we see the opposite trend, with centrals being quenchier than satellites at fixed $M_{*}$, sSFR, and $\Sigma$ bins in an average sense out to $z$ $\sim$ 1. 

Wet mergers can boost star-formation due to gas compression in short timescales, followed by subsequent quenching over a longer timescale due to gas consumption and the potential rejuvenation of the nuclear activity in merging systems (e.g.; \citealp{Mihos96,Ellison08,Ellison13}). There is also evidence for the merger rate being higher for more massive systems (e.g.; \citealp{Patton08,Xu12,Robotham14}). Therefore, the merger scenario can potentially explain the fast bursting of star-formation for massive, high sSFR star-forming galaxies. The merger picture might also explain why centrals are quenching faster than satellites, if the dominant wet mergers in centrals had happened much earlier than satellites so that at the present, we are mainly witnessing the quenching phase of the merger. An even older ($>$ 600 Myr) star-formation derivative might reveal the past bursting phase of centrals. Moreover, whether the wet merger is ``major'' or ``minor'' might also explain why centrals are quenching faster than satellites. In other words, the current bursting of star-formation in massive satellites might be because of wet major mergers, whereas this recent bursting in massive centrals could be mostly due to wet minor mergers.   

Although the wet merger scenario might explain the burstiness of massive galaxies in the field as seen in e.g., Figure \ref{fig:2D-MD} \citep{Lin10}, the merger picture becomes problematic in denser environments. Mergers are more common in denser environments than the field (especially in group-scales; \citealp{Perez09,Sobral11,Tonnesen12}) but we see in e.g. Figures \ref{fig:mass-density-SFA-SFJ} and \ref{fig:hz-mass-density-SFA-SFJ} that satellites and the overall galaxy distribution follow similar bursting trends. Moreover, there is evidence that centrals are as old as or even younger than satellites and they are quenched at the same time as or even more recently than satellites of the same mass \citep{Pasquali10,Fitzpatrick15,Smethurst17}. Moreover, mergers in dense environments seem to be mostly dry and the merger-driven history of centrals is mainly due to gas-poor, dry mergers with no significant star-formation (e.g.; \citealp{McIntosh08,Lin10,Lidman12,Shankar15,Davidzon16}). 

By studying SDSS galaxy pairs \cite{Ellison10} showed that although interactions happen at all environments, interaction-triggered star-formation is seen only in low-to-intermediate density environments. The position of centrals in the densest regions of groups/clusters is in agreement with this picture and explains why they avoid bursting especially in densest regions as clearly seen in e.g., Figure \ref{fig:2D-MD}. 

There is evidence for the gas fraction being higher in the field galaxies than those in denser environments as previous accretion of galaxies into their current halos made them gas-stripped \citep{Cortese09,Fabello12,Catinella13,Boselli14}. It is also possible that the cold gas accretion from the surrounding LSS can more easily/efficiently/numerously penetrate isolated and satellite galaxies than centrals that are located in the densest regions of groups/clusters. \cite{vandeVoort17} simulations show that more massive centrals and satellites both have a higher gas accretion rate than less massive ones (also see \citealp{Dekel13}). They also find that gas accretion rate is lower or fully suppressed (depending on the halo mass) in the center of halos. They also find a strong environmental dependence of accretion rate primarily for satellites. This might explain the strong quenching of low-mass satellites in denser regions seen in Figure \ref{fig:2D-MD}. Therefore, a combination of stellar mass and halo-centric and environmental dependence of gas accretion rate, gas fraction, and mergers can potentially explain the observed trends. Further analyses including gas and age dependence of different galaxy types on SFA, SFJ, local environment, stellar mass, and sSFR can shed light on this.

The recent mass quenching of lower mass galaxies (log($M_{*}/M_{\odot}$) $\lesssim$ 10) might also be partly due to fast-acting stellar and supernova feedback which is stronger on less massive systems because of their weaker gravitational potential (e.g.; \citealp{Ceverino09,Hopkins14}). However, feedback produces short-timescale burstiness in star formation especially for less massive galaxies (e.g.; \citealp{Hopkins14}), which is not clearly seen in the median trends in our results. However, part of this discrepancy might be due to different rates for star-formation burstiness and quenching. That is, when a galaxy bursts, it almost instantly becomes blue but the quenching phase takes longer to act. Therefore, the average in bins of $M_{*}$-sSFR would be quenching because of this asymmetry. Rapid AGN feedback on less massive galaxies can also have some effect in their quenching \citep{Smethurst17}.  

The higher quenching of our high-$z$ sample compared to SDSS, even at fixed $M_{*}$, sSFR, and environment might be due to a faster and more efficient quenching rate at higher redshifts. By studying a large sample of transiting galaxies at $z$ $\sim$ 1, \cite{goncalves12} showed that mass flux from the blue cloud to the red sequence (through the green valley) is larger at $z$ $\sim$ 1 compared to the same quantity in the local universe \citep{Martin07}. The faster quenching rate of transiting galaxies at $z$ $\sim$ 1 has been recently reconfirmed by \cite{Cavalcante18}. In this paper, however, we find that this seems to be true for all galaxies (not just transiting systems) and even satellites and centrals. Quenching, regardless of the process, is stronger at $z$ $\sim$ 1 for all galaxies at all stellar masses, sSFRs, and $\Sigma$s. In agreement with our results, \cite{Tinker10} and \cite{Quadri12} showed that satellite quenching must proceed faster at high redshift. Moreover, \cite{McGee14} argued that given the strong redshift evolution of the star-formation rate, the quenching timescales should be shorter at higher redshift. Recently, \cite{Rowlands18} also found a faster quenching at $z$ $\sim$ 0.7 than the local universe.

Out to $z$ $\sim$ 1, the local environment seems to have a milder effect (if any) on the SFA and SFJ, with stellar mass (and sSFR) being the dominant factor over a large range of $M_{*}$, $\Sigma$, sSFRs and redshift. The dominance of ``mass quenching'' over ``environmental quenching'' has been found, particularly for more massive galaxies and at higher redshifts (e.g.; \citealp{Peng10,Darvish16,Smethurst17}). Only very dense environments seem to significantly influence the SFA, particularly for the least massive satellites. Ram pressure stripping and other environmentally-driven processes that are practically effective on less massive galaxies might be causing this, whereas mergers are likely behind the trends for more massive galaxies in the field. 

The milder (lack of) environmental effects on SFA and SFJ might also be due to the typical long timescales of environmental quenching due to e.g. strangulation (e.g.; \citealp{Balogh00,Peng15}). However, there is no agreement on the timescales either, as some studies found a short environmental quenching timescale, particularly for less massive satellites (e.g.; \citealp{Boselli08,Peng10,Darvish17,Crossett17}). In addition, simulations of group and cluster galaxies by \cite{Bahe15} show that ram pressure stripping is more affective at $z$ $\sim$ 1 than $z$=0, in agreement with a higher SFA we find for our high-$z$ sample compared to that of the local-universe at a fixed local environment. We further note that our results indicate that mass quenching likely happens in short timescales, while the opposite is likely true for environmental quenching (at least when averaged over the whole population). That threshold, established by our methodology (especially given the lack of the SFJ at slightly longer timescales) might set an interesting comparison basis for future studies. Further studies using a modified SFA measure derived over a long timescale are needed to investigate this.     
                
Using \cite{Martin17} method, in the future, we will use other physical parameters such as age, gas mass, and gas/stellar metallicity and define new parameters (such as derivatives of SFA over different timescales) to resolve some of the issues stated here and also perform new studies that would potentially shed light on the physics of quenching and bursting in galaxies. Moreover, complementary to the LEGA-C survey, high S/N continuum spectroscopy of (particularly) low-mass galaxies (log($M_{*}$/$M_{\odot}$) $\lesssim$ 10) located in very dense environments (log$\Sigma$ $\gtrsim$ 1) at $z$ $\sim$ 1 is essential to further investigate the potential environmental trends we already found in the local universe.       

\section{summary} \label{sum}

We study the ``quenching'' and ``bursting'' of galaxies as a function of stellar mass ($M_{*}$), local environment ($\Sigma$), and specific star-formation rate (sSFR) using $\sim$ 123,000 $GALEX$/SDSS galaxies at $z$ $\approx$ 0.02-0.12 and $\sim$ 420 $GALEX$/COSMOS/LEGA-C galaxies at $z$ $\approx$ 0.6-1.0 with high S/N continuum spectra. To quantify recent quenching and bursting of galaxies, we define the star formation acceleration (SFA) and the star formation jerk (SFJ) presented as the time derivative of the extinction-corrected $NUV-i$ color over the past 300 Myr and 600-300 Myr, respectively \citep{Martin17}. The key results from this work are as follows:

\begin{enumerate}

\item To $z$ $\sim$ 1 and at fixed sSFR and $\Sigma$, on average, less massive galaxies are quenching, whereas more massive systems are bursting, with a quenching/bursting transition at log($M_{*}$/$M_{\odot}$) $\sim$ 10.5-11 and likely a short quenching/bursting timescale ($\lesssim$ 300 Myr).

\item The bursting of star-formation happens mostly in massive (log($M_{*}$/$M_{\odot}$) $\gtrsim$ 11), high sSFR galaxies (log(sSFR/Gyr$^{-1}$) $\gtrsim$ -2), particularly those in the field (log($\Sigma$/Mpc$^{-2}$) $\lesssim$ 0; and among group galaxies, satellites more than centrals). 

\item Most of the quenching of star-formation happens in low-mass (log($M_{*}$/$M_{\odot}$) $\lesssim$ 9), low sSFR galaxies (log(sSFR/Gyr$^{-1}$) $\lesssim$ -2), in particular those located in dense environments (log($\Sigma$/Mpc$^{-2}$) $\gtrsim$ 1). For central galaxies, quenching is significant even for massive systems. These show the combined effects of $M_{*}$ and $\Sigma$ in quenching/bursting of galaxies since $z$ $\sim$ 1.

\item Stellar mass seems to have stronger effects than local environment on recent quenching/bursting of galaxies to $z$ $\sim$ 1.

\item The strength of mass and environmental quenching/bursting is larger at higher redshift. Quenching, regardless of its nature, is stronger at higher redshifts.

\item Among group galaxies, and at any given $M_{*}$, sSFR, and $\Sigma$, centrals are quenchier (quenching faster) than satellites in an average sense. 

\end{enumerate}

Since the LEGA-C survey is not designed \textit{a priori} to target galaxies in dense environments, then complementary to the LEGA-C survey, high S/N continuum spectroscopy of (particularly) low-mass galaxies (log($M_{*}$/$M_{\odot}$) $\lesssim$ 10) located in very dense environments (log$\Sigma$ $\gtrsim$ 1) at $z$ $\sim$ 1 is necessary to further investigate and more robustly constrain the potential environmental trends we already found in this work. 

\section*{acknowledgements}

We are immensely grateful to the anonymous referee for reading the manuscript and providing very useful comments that improved the quality of this paper. We would like to gratefully thank the LEGA-C team for providing the catalog of spectral indices which was used to compare with the extracted indices for sanity checks. B.D. acknowledges financial support from NASA through the Astrophysics Data Analysis Program (ADAP), grant number NNX12AE20G, and the National Science Foundation, grant number 1716907. B.D. is grateful to Alessandro Rettura for the thoughtful discussions and Mark Seibert for providing further information regarding the available $GALEX$ data. D.S. acknowledges financial support from the Netherlands Organisation for Scientific research (NWO) through a Veni fellowship and from Lancaster University through an Early Career Internal Grant A100679. Based on data products from observations made with ESO Telescopes at the La Silla Paranal Observatory under program ID 194.$A$-2005($A$-$F$). $GALEX$ (Galaxy Evolution Explorer) is a NASA Small Explorer, launched in April 2003. We gratefully acknowledge NASA's support for construction, operation, and science analysis for the $GALEX$ mission, developed in cooperation with the Centre National d'Etudes Spatiales of France and the Korean Ministry of Science and Technology. Funding for SDSS-III has been provided by the Alfred P. Sloan Foundation, the Participating Institutions, the National Science Foundation, and the U.S. Department of Energy Office of Science. The SDSS-III web site is http://www.sdss3.org/. SDSS-III is managed by the Astrophysical Research Consortium for the Participating Institutions of the SDSS-III Collaboration including the University of Arizona, the Brazilian Participation Group, Brookhaven National Laboratory, Carnegie Mellon University, University of Florida, the French Participation Group, the German Participation Group, Harvard University, the Instituto de Astrofisica de Canarias, the Michigan State/Notre Dame/JINA Participation Group, Johns Hopkins University, Lawrence Berkeley National Laboratory, Max Planck Institute for Astrophysics, Max Planck Institute for Extraterrestrial Physics, New Mexico State University, New York University, Ohio State University, Pennsylvania State University, University of Portsmouth, Princeton University, the Spanish Participation Group, University of Tokyo, University of Utah, Vanderbilt University, University of Virginia, University of Washington, and Yale University. 

\appendix

\section{A. Completeness Corrections for the Local-universe Sample} \label{A}

The SDSS spectroscopic sample has an incompleteness associated with magnitude and the mechanical restrictions due to fiber collision that does not allow to obtain redshifts for all the galaxies that are closer than 55$\arcsec$ on the sky. This results in significant spectroscopic incompleteness in crowded regions of the sky such as galaxy clusters and groups that are very important in this study. We estimate the magnitude and fiber collision related completeness by comparing our primary spectroscopic sample (sample A) with the main galaxy sample used to select targets for spectroscopy \citep{Strauss02}. Similar to \cite{Baldry06}, we divide galaxies in the main galaxy sample into classes based on their magnitude and the number of neighbors within 55$\arcsec$ (zero, one, and greater than one). The spectroscopic completeness, $C$, for a class is determined by the number of available spectra divided by the total number of objects in each class. 

\begin{figure}
\centering
\includegraphics[width=3.5in]{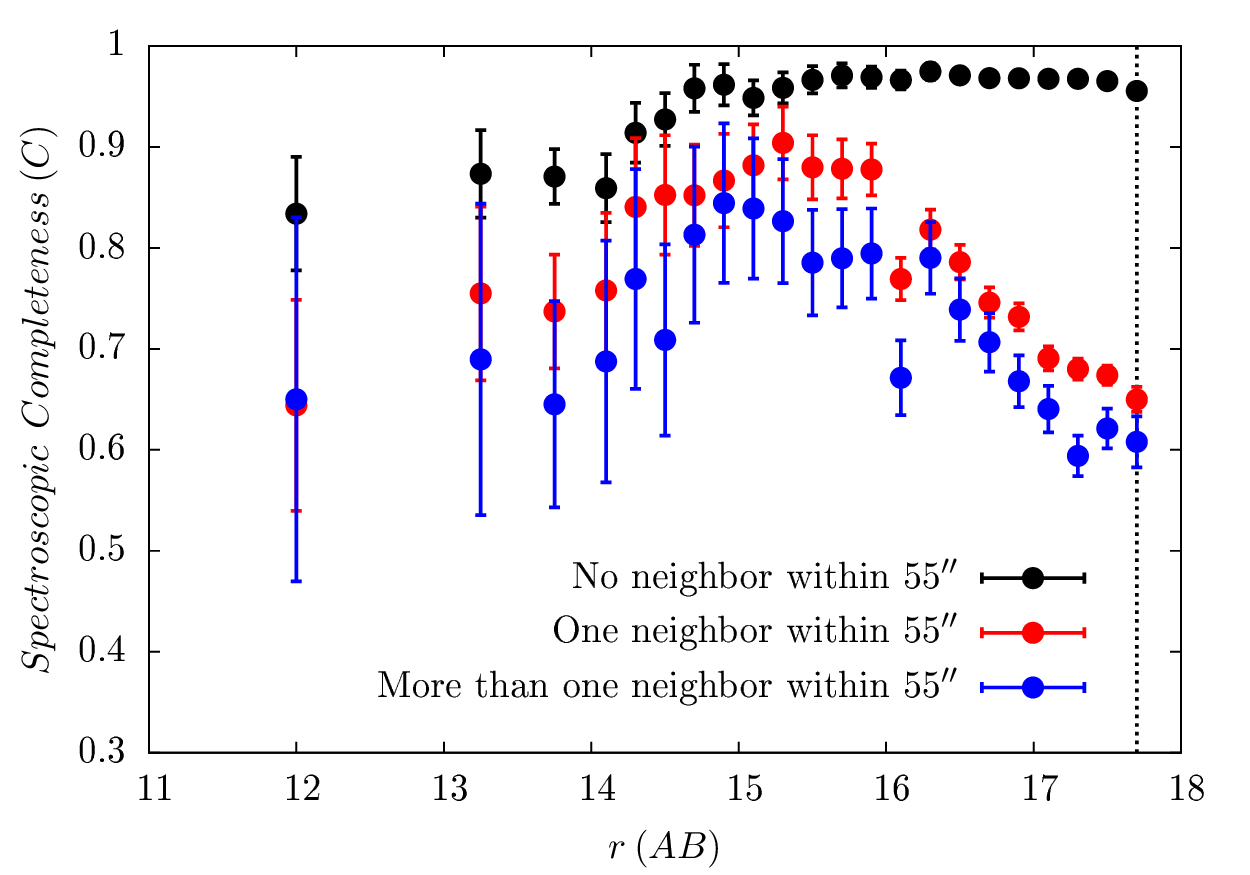}
\caption{Spectroscopic completeness as a function of Petrosian magnitude $r$ and for galaxies with zero, one, and greater than one neighbors within 55$\arcsec$. The vertical dashed line shows the magnitude limit used for sample selection ($r$ $\leqslant$ 17.7). Beyond this, spectroscopic completeness decreases significantly. The uncertainties are poissonian.}
\label{fig:comp-den}
\end{figure}

Figure \ref{fig:comp-den} shows the completeness as a function of Petrosian magnitude $r$ and for galaxies with zero, one, and greater than one neighbors within 55$\arcsec$. As expected, the spectroscopic completeness is higher for galaxies with no neighbors within 55$\arcsec$ and this comprises the majority of our spectroscopic sample ($\sim$ 89\%). Only $\sim$ 9\% and $\sim$ 2\% of our main spectroscopic sample includes galaxies with one and greater than one neighbor within 55$\arcsec$, respectively. The spectroscopic sample is $>$ 95 \% complete at $r$ $\gtrsim$ 15.3 for galaxies with no neighbors within 55$\arcsec$ and that comprises the majority of our primary sample ($\sim$ 86\%). We weight the estimated local density of galaxies by 1/$C$ to take the spectroscopic incompleteness due to magnitude and fiber collision into account. 

\begin{figure}
\centering
\includegraphics[width=3.5in]{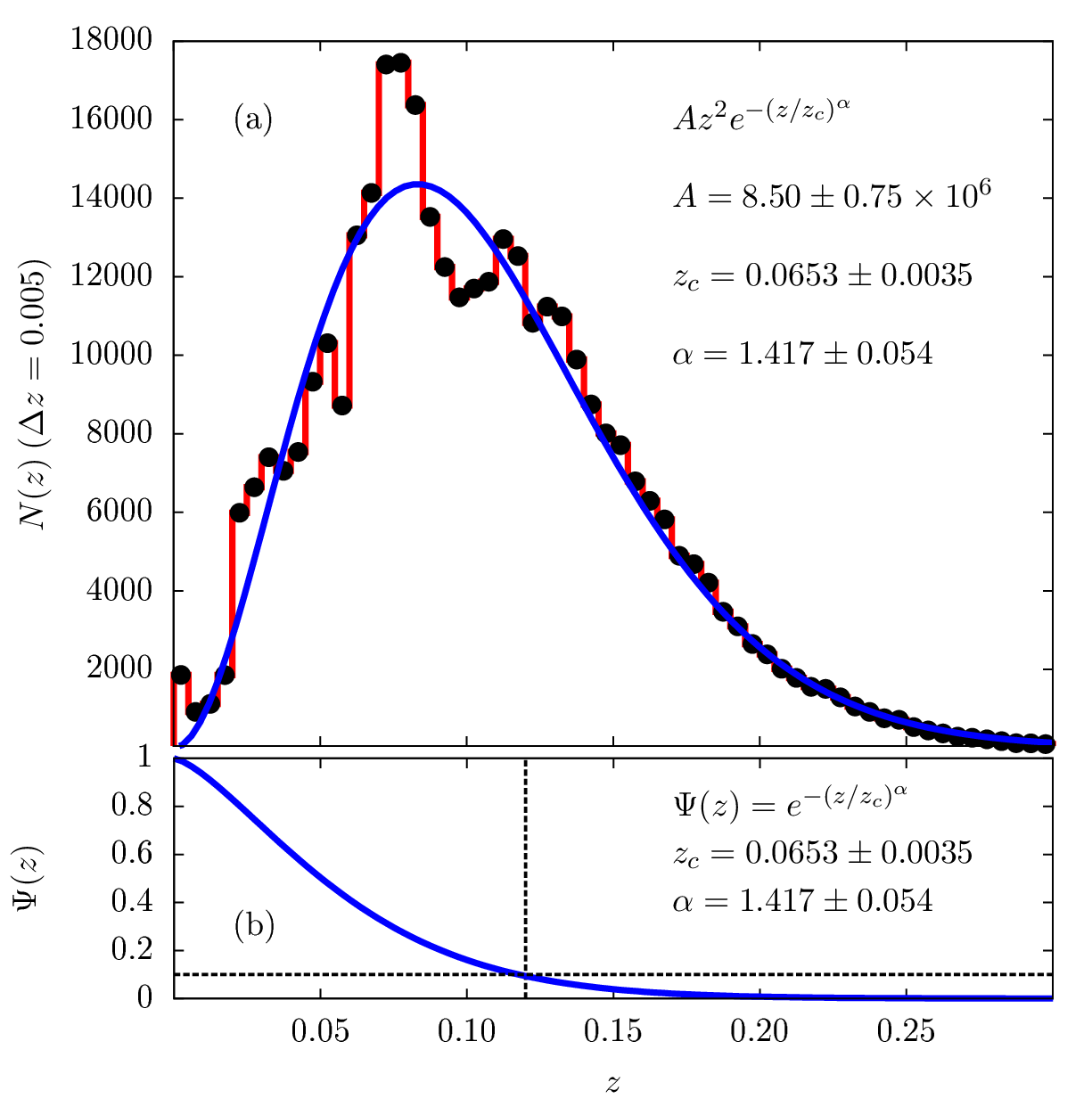}
\caption{(a) Spectroscopic redshift distribution of the local-universe data (in bins of $\Delta z$=0.005; red histogram) and the best fitted model ($z^{2}e^{-(z/z_{c})^{\alpha}}$; blue solid curve) given by $A$=8.50$\pm$0.75 $\times$ 10$^{6}$, $z_{c}$=0.0653$\pm$0.0035, and $\alpha$=1.417$\pm$0.054. (b) Selection function ($\Psi$) as a function of redshift. Horizontal dashed line shows where $\Psi$=0.1. This approximately corresponds to $z$=0.12, indicated by the vertical dashed line.}
\label{fig:sf}
\end{figure}

Since we use a flux-limited spectroscopic sample ($r$ $\leqslant$ 17.7), we need to compensate for the decrease in the number density of galaxies with increasing redshift when estimating the local density of galaxies. Otherwise, we unrealistically underestimate the local density of galaxies at higher redshifts. As we explained in Section \ref{method-env-sdss}, we model the change in the mean number of galaxies (in redshift bins of $\Delta z$=0.005) as a function of redshift with:
\begin{equation} 
N(z)dz=Az^{2}e^{-(z/z_{c})^{\alpha}}dz
\end{equation}
According to equation A1, the mean number of galaxies increases with increasing redshift as $z^{2}$ as we cover a larger spatial volume and at the same time, it decreases with the selection function ($\Psi$= $e^{-(z/z_{c})^{\alpha}}$) since the intrinsically fainter galaxies are no longer observable at higher redshifts. Figure \ref{fig:sf} (a) shows the redshift distribution of the data and the best fitted model given by $A$=8.50$\pm$0.75 $\times$ 10$^{6}$, $z_{c}$=0.0653$\pm$0.0035, and $\alpha$=1.417$\pm$0.054. We correct the local density of each galaxy by a weight 1/$\Psi(z)$. Figure \ref{fig:sf} (b) shows the selection function as a function of redshift. In order to avoid large uncertainties and fluctuations in the estimated densities due to a smaller sample size at higher redshifts, we only use galaxies for which $\Psi(z)$ $\geqslant$ 0.1. This corresponds to $z$ $\sim$ 0.12. We also define a lower redshift cut of $z$=0.02 to avoid issues related to local motions, edge effects, and the photometry of very bright, nearby sources.

\begin{figure*}
\centering
\includegraphics[width=7in]{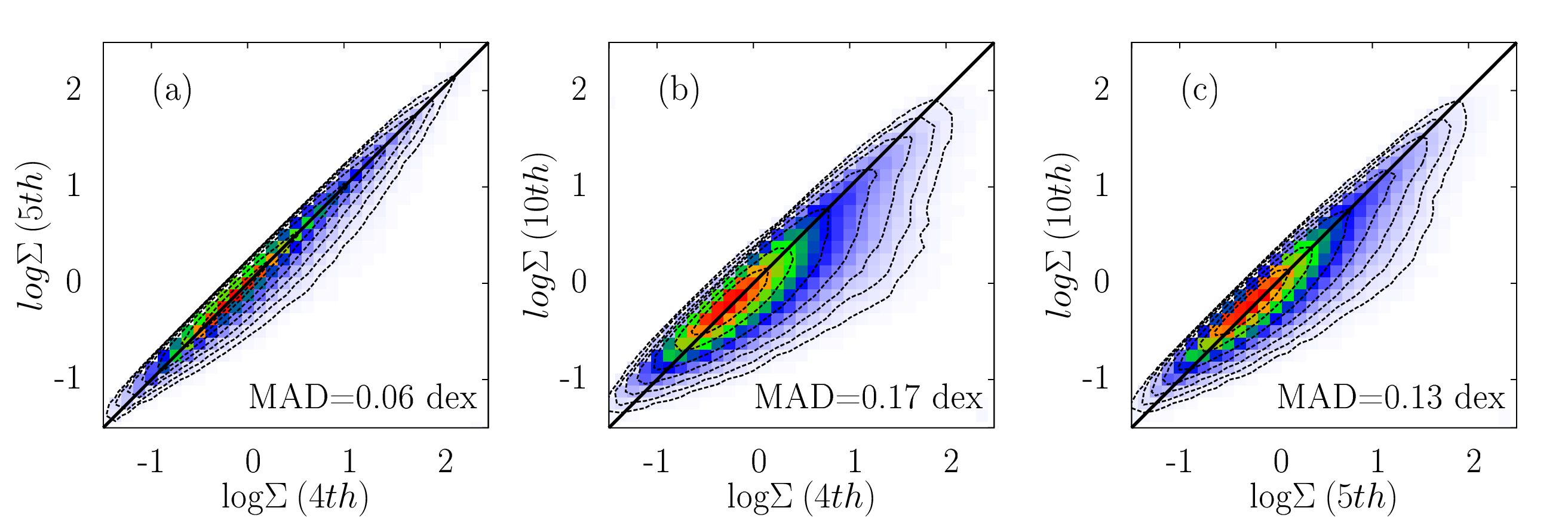}
\caption{Comparison between the estimated surface densities using $4th$ and $5th$ (a), $4th$ and $10th$ (b), and $5th$ and $10th$ (c) nearest neighbor methods. The median absolute deviation (MAD) between the density values is also shown. There is an overall good agreement between the estimated densities, with only 0.3\% ($4th$ and $5th$), 2.7\% ($4th$ and $10th$), and 1.2\% ($5th$ and $10th$) of the sources are more than 1 dex different in densities.}
\label{fig:comparison}
\end{figure*}

The performance of the nearest neighbor method used in Section \ref{method-env-sdss} depends on the selection of the value $N$ ($nth$ nearest neighbor) as discussed in \cite{Darvish15a}. A small value may result in unrealistically large local densities because of Poisson noise and random clustering of spatially uncorrelated galaxies, whereas a large value tends to oversmooth the details of galaxy distribution and is prone to underestimation of local densities. We compare estimated surface densities using $4th$, $5th$, and $10th$ nearest neighbor methods as shown in Figure \ref{fig:comparison}. The median absolute deviation between density values of $4th$ and $5th$, $4th$ and $10th$, and $5th$ and $10th$ methods are 0.06, 0.17, and 0.13 dex, respectively. Only 0.3, 2.7, and 1.2\% of the sources are more than 1 dex different in density values when we compare $4th$ and $5th$, $4th$ and $10th$, and $5th$ and $10th$ estimations, respectively. However, we note that the density estimations are slightly biased towards higher values for lower $nth$ estimators due to the nature of the nearest neighbor method. \cite{Darvish15a} performed two sets of simulations, comparing the performance of different density estimators. Both sets of simulations show that the $10th$ nearest neighbor outperforms the $5th$ nearest neighbor method. Hence, we use $N$=10 in Section \ref{method-env-sdss}.

Another parameter that affects the estimated surface densities in Section \ref{method-env-sdss} is $\Delta v$. A small value of $\Delta v$ underestimates the surface densities in dense regions due to the finger-of-god effect, whereas a large value affects the low surface densities. Here, we use $\Delta v$=$\pm$1000 kms$^{-1}$ which is equivalent to the typical radial velocity dispersion of rich galaxy clusters and is large enough to suppress the finger-of-god effect. Moreover, \cite{Cooper05} showed that a velocity range of $\pm$1000-1500 kms$^{-1}$ is best suited for environmental studies in a broad range of environments. 

\section{B. Adaptive Kernel vs. $10th$ Nearest Neighbor} \label{B}

\begin{figure}
\centering
\includegraphics[width=3.5in]{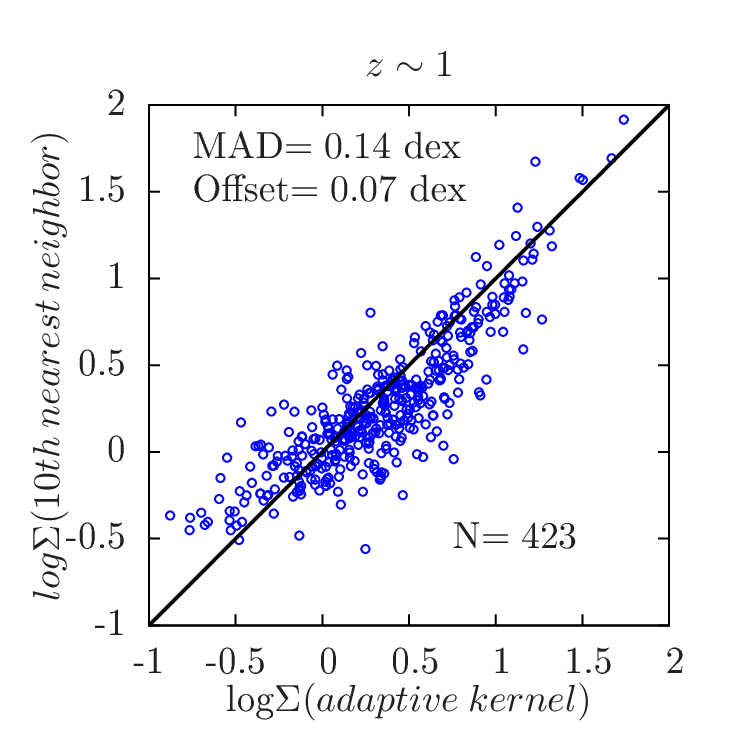}
\caption{Comparison between the estimated surface densities using the distance to the $10th$ nearest neighbor and the adaptive kernel smoothing methods for our high-$z$ sample. The median offset and the median absolute deviation (MAD) between the density values are also shown. We find an overall good agreement between the two methods.}
\label{fig:comp-knn}
\end{figure}

Using different density estimators in the local universe (projected distance to the $10th$ nearest neighbor) and high-$z$ (adaptive kernel smoothing) might lead to a bias when we compare the results at low and high redshift. To investigate this, we perform the density estimation using the $10th$ nearest neighbor method for our high-$z$ sample as well and compare it with that of the adaptive kernel smoothing. Figure \ref{fig:comp-knn} shows the comparison. We find a relatively good agreement between the two, with a median offset of $\sim$ 0.07 dex (log$\Sigma$(Kernel)-log$\Sigma$($10th$)) and a median absolute deviation of $\sim$ 0.14 dex. Therefore, the selection of different density estimators does not have significant effects in the presented results.

\begin{figure*}
\centering
\includegraphics[width=7.0in]{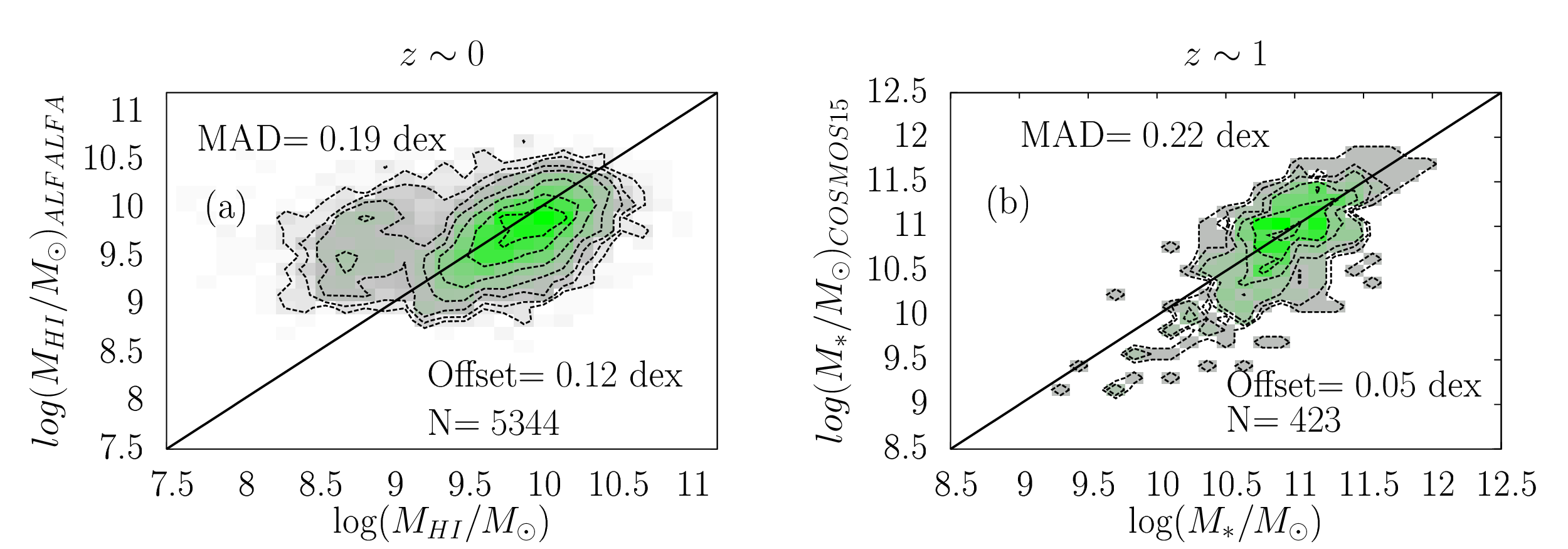}
\caption{(a) Comparison between the atomic hydrogen mass $M_{HI}$ measurements from the ALFALFA survey \citep{Haynes11} and estimations based on \cite{Martin17} methodology for 5344 sources matched with our local-universe sample. There is a good agreement between the two at log($M_{HI}/M_{\odot}$) $\gtrsim$ 9, with an offset and the median absolute deviation of $\sim$ 0.1 and 0.2 dex, respectively. (b) Comparison between the stellar masses from the COSMOS15 catalog \citep{Laigle16} and those estimated for our high-$z$ LEGA-C sample at $z$ $\sim$ 1 using \cite{Martin17} method. Only a small offset of $\sim$ 0.05 and a median absolute deviation of $\sim$ 0.2 is seen between the two.}
\label{fig:more-comp}
\end{figure*}
   
\section{C. Comparison} \label{C}

\cite{Martin17} performed a comparison between some physical parameters and trends based on their method and those in the literature at $z$ $\sim$ 0 and found a relatively good agreement, with small off-sets and deviations. Here, we present two more comparisons, highlighting the robustness and reliability of the method.

Figure \ref{fig:more-comp} (a) shows the comparison between the atomic hydrogen mass ($M_{HI}$) measurements from the ALFALFA survey \citep{Haynes11} and estimations based on \cite{Martin17} methodology for 5344 sources matched with our local-universe sample. Note that \cite{Martin17} gives the total cold gas mass including atomic hydrogen ($HI$), molecular hydrogen ($H_{2}$), helium ($He$), and a small fraction of metals. We convert the total gas mass to $M_{HI}$ assuming that the composition of the cold gas in the local universe is 59\% $HI$, 15\% $H_{2}$, and 26\% $He$ and metals \citep{Obreschkow09}. According to Figure \ref{fig:more-comp} (a), for log($M_{HI}/M_{\odot}$) $\gtrsim$ 9, there is a good agreement between the two, with a small offset of $\sim$ 0.1 dex and the median absolute deviation of $\sim$ 0.2 dex. Note that part of the dispersion might be due to the mass and galaxy-type dependence of $M_{H_{2}}$/$M_{HI}$ ratio. The cause of the offset and the disagreement at low-masses (log($M_{HI}/M_{\odot}$) $\lesssim$ 9) is beyond the scope of this work. 

Figure \ref{fig:more-comp} (b) compares the stellar masses from the COSMOS15 catalog \citep{Laigle16} and those estimated for our high-$z$ LEGA-C sample at $z$ $\sim$ 1 using \cite{Martin17} method. Only a small offset of $\sim$ 0.05 and a median absolute deviation of $\sim$ 0.2 is seen between the two, indicating that the methodology used here gives reasonable physical values even at higher redshifts.        

\begin{figure*}
\centering
\includegraphics[width=7.0in]{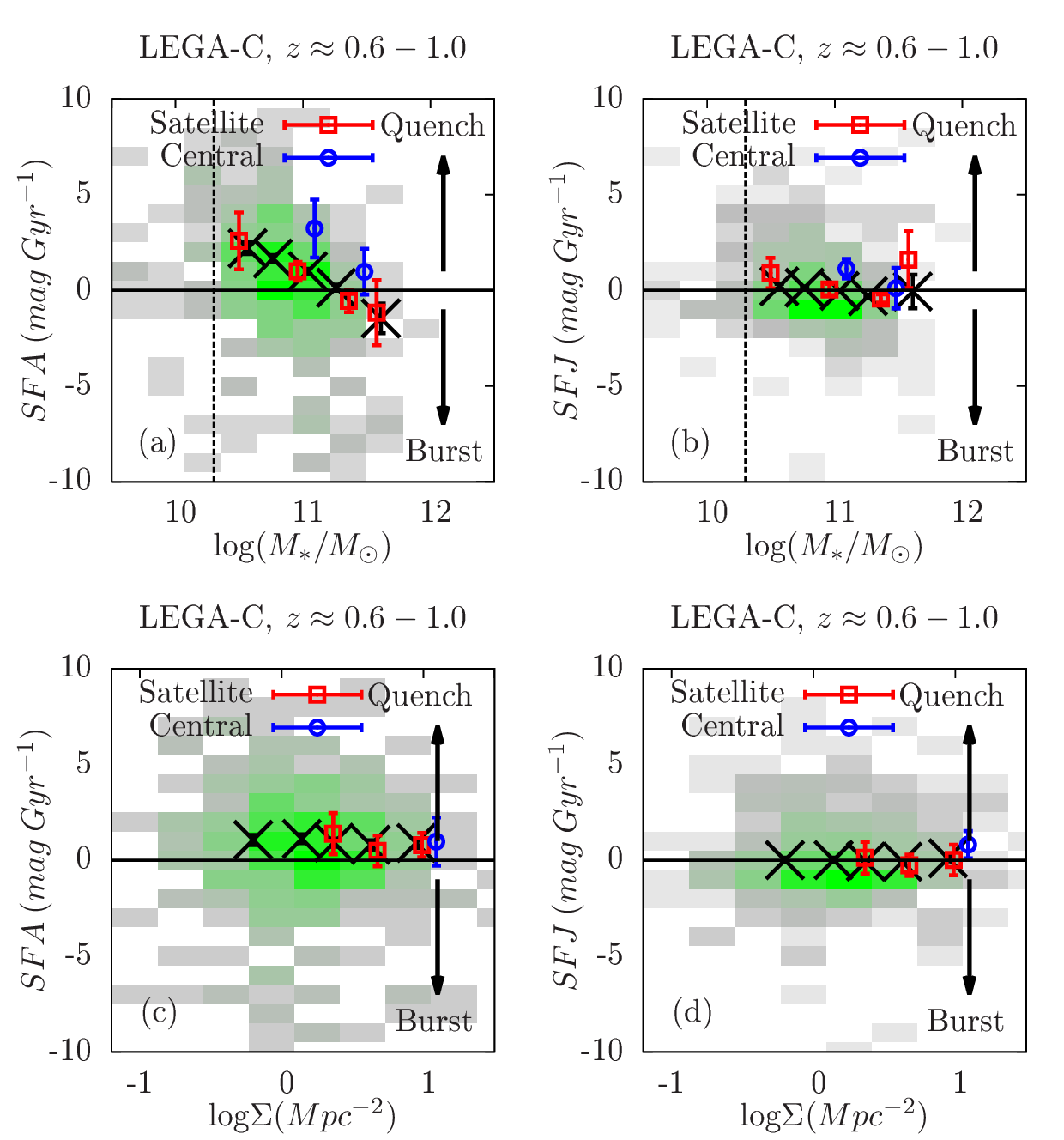}
\caption{Similar to Figure \ref{fig:hz-mass-density-SFA-SFJ} for the LEGA-C high-$z$ sample at $z$ $\sim$ 1 but for group galaxies whose groups have $>$ 10 members. Red, blue, and black points show the median values for satellite, central, and all galaxies. Similar to Figure \ref{fig:hz-mass-density-SFA-SFJ}, we can still retrieve the trends between SFA (or SFJ), $M_{*}$, and $\Sigma$ and for satellites and centrals.}
\label{fig:hz-mass-density-SFA-SFJ10}
\end{figure*}

\section{D. High-$z$ Results Using Groups with $>$ 10 Members} \label{D}

In presenting the high-$z$ sample results, in order to increase the sample size and a broader range of the physical parameters involved ($M_{*}$, $\Sigma$, and sSFR), all groups with $\geqslant$ 2 members are considered. This is different than our low-$z$ sample selection where groups with $>$ 10 members are considered. In addition to potential biases that might arise when comparing the low and high redshift samples, this selection of high-$z$ sample (all groups with $\geqslant$ 2 members) makes it more vulnerable to contamination by interlopers. We check whether we can still recover the global trends between SFA (or SFJ), $M_{*}$, and $\Sigma$ for our high-$z$ sample by choosing all group galaxies whose groups have $>$ 10 members. This leads to a factor $\sim$ 2 and $\sim$ 9 reduction in the number of satellites and centrals, respectively. However, as shown in Figure \ref{fig:hz-mass-density-SFA-SFJ10}, even with this limited sample size, we can retrieve the trends that are already shown in Section \ref{science-highz} and Figure \ref{fig:hz-mass-density-SFA-SFJ}. Therefore, the presented results are likely not much affected by selecting $\geqslant$ 2 group membership for our high-$z$ sample.

\bibliographystyle{aasjournal} 
\bibliography{references}

\begin{thebibliography}{}
\expandafter\ifx\csname natexlab\endcsname\relax\def\natexlab#1{#1}\fi
\providecommand{\url}[1]{\href{#1}{#1}}

\bibitem[{{Alam} {et~al.}(2015){Alam}, {Albareti}, {Allende Prieto}, {Anders},
  {Anderson}, {Anderton}, {Andrews}, {Armengaud}, {Aubourg}, {Bailey}, \&
  et~al.}]{Alam15}
{Alam}, S., {Albareti}, F.~D., {Allende Prieto}, C., {et~al.} 2015, \apjs, 219,
  12

\bibitem[{{Bah{\'e}} \& {McCarthy}(2015)}]{Bahe15}
{Bah{\'e}}, Y.~M., \& {McCarthy}, I.~G. 2015, \mnras, 447, 969

\bibitem[{{Baldry} {et~al.}(2006){Baldry}, {Balogh}, {Bower}, {Glazebrook},
  {Nichol}, {Bamford}, \& {Budavari}}]{Baldry06}
{Baldry}, I.~K., {Balogh}, M.~L., {Bower}, R.~G., {et~al.} 2006, \mnras, 373,
  469

\bibitem[{{Balogh} {et~al.}(1999){Balogh}, {Morris}, {Yee}, {Carlberg}, \&
  {Ellingson}}]{Balogh99}
{Balogh}, M.~L., {Morris}, S.~L., {Yee}, H.~K.~C., {Carlberg}, R.~G., \&
  {Ellingson}, E. 1999, \apj, 527, 54

\bibitem[{{Balogh} {et~al.}(2000){Balogh}, {Navarro}, \& {Morris}}]{Balogh00}
{Balogh}, M.~L., {Navarro}, J.~F., \& {Morris}, S.~L. 2000, \apj, 540, 113

\bibitem[{{Bekki} \& {Couch}(2003)}]{Bekki03}
{Bekki}, K., \& {Couch}, W.~J. 2003, \apjl, 596, L13

\bibitem[{{Berti} {et~al.}(2017){Berti}, {Coil}, {Behroozi}, {Eisenstein},
  {Bray}, {Cool}, \& {Moustakas}}]{Berti17}
{Berti}, A.~M., {Coil}, A.~L., {Behroozi}, P.~S., {et~al.} 2017, \apj, 834, 87

\bibitem[{{Best} {et~al.}(2005){Best}, {Kauffmann}, {Heckman}, {Brinchmann},
  {Charlot}, {Ivezi{\'c}}, \& {White}}]{Best05}
{Best}, P.~N., {Kauffmann}, G., {Heckman}, T.~M., {et~al.} 2005, \mnras, 362,
  25

\bibitem[{{Birnboim} \& {Dekel}(2003)}]{Birnboim03}
{Birnboim}, Y., \& {Dekel}, A. 2003, \mnras, 345, 349

\bibitem[{{Bluck} {et~al.}(2014){Bluck}, {Mendel}, {Ellison}, {Moreno},
  {Simard}, {Patton}, \& {Starkenburg}}]{Bluck14}
{Bluck}, A.~F.~L., {Mendel}, J.~T., {Ellison}, S.~L., {et~al.} 2014, \mnras,
  441, 599

\bibitem[{{Boselli} {et~al.}(2008){Boselli}, {Boissier}, {Cortese}, \&
  {Gavazzi}}]{Boselli08}
{Boselli}, A., {Boissier}, S., {Cortese}, L., \& {Gavazzi}, G. 2008, \apj, 674,
  742

\bibitem[{{Boselli} \& {Gavazzi}(2006)}]{Boselli06}
{Boselli}, A., \& {Gavazzi}, G. 2006, \pasp, 118, 517

\bibitem[{{Boselli} \& {Gavazzi}(2014)}]{Boselli14}
---. 2014, \aapr, 22, 74

\bibitem[{{Bruzual} \& {Charlot}(2003)}]{Bruzual03}
{Bruzual}, G., \& {Charlot}, S. 2003, \mnras, 344, 1000

\bibitem[{{Capak} {et~al.}(2007){Capak}, {Aussel}, {Ajiki}, {McCracken},
  {Mobasher}, {Scoville}, {Shopbell}, {Taniguchi}, {Thompson}, {Tribiano},
  {Sasaki}, {Blain}, {Brusa}, {Carilli}, {Comastri}, {Carollo}, {Cassata},
  {Colbert}, {Ellis}, {Elvis}, {Giavalisco}, {Green}, {Guzzo}, {Hasinger},
  {Ilbert}, {Impey}, {Jahnke}, {Kartaltepe}, {Kneib}, {Koda}, {Koekemoer},
  {Komiyama}, {Leauthaud}, {Le Fevre}, {Lilly}, {Liu}, {Massey}, {Miyazaki},
  {Murayama}, {Nagao}, {Peacock}, {Pickles}, {Porciani}, {Renzini}, {Rhodes},
  {Rich}, {Salvato}, {Sanders}, {Scarlata}, {Schiminovich}, {Schinnerer},
  {Scodeggio}, {Sheth}, {Shioya}, {Tasca}, {Taylor}, {Yan}, \&
  {Zamorani}}]{Capak07}
{Capak}, P., {Aussel}, H., {Ajiki}, M., {et~al.} 2007, \apjs, 172, 99

\bibitem[{{Catinella} {et~al.}(2013){Catinella}, {Schiminovich}, {Cortese},
  {Fabello}, {Hummels}, {Moran}, {Lemonias}, {Cooper}, {Wu}, {Heckman}, \&
  {Wang}}]{Catinella13}
{Catinella}, B., {Schiminovich}, D., {Cortese}, L., {et~al.} 2013, \mnras, 436,
  34

\bibitem[{{Ceverino} \& {Klypin}(2009)}]{Ceverino09}
{Ceverino}, D., \& {Klypin}, A. 2009, \apj, 695, 292

\bibitem[{{Chilingarian} {et~al.}(2010){Chilingarian}, {Melchior}, \&
  {Zolotukhin}}]{Chilingarian10}
{Chilingarian}, I.~V., {Melchior}, A.-L., \& {Zolotukhin}, I.~Y. 2010, \mnras,
  405, 1409

\bibitem[{{Chilingarian} \& {Zolotukhin}(2012)}]{Chilingarian12}
{Chilingarian}, I.~V., \& {Zolotukhin}, I.~Y. 2012, \mnras, 419, 1727

\bibitem[{{Cooper} {et~al.}(2005){Cooper}, {Newman}, {Madgwick}, {Gerke},
  {Yan}, \& {Davis}}]{Cooper05}
{Cooper}, M.~C., {Newman}, J.~A., {Madgwick}, D.~S., {et~al.} 2005, \apj, 634,
  833

\bibitem[{{Cortese} \& {Hughes}(2009)}]{Cortese09}
{Cortese}, L., \& {Hughes}, T.~M. 2009, \mnras, 400, 1225

\bibitem[{{Crossett} {et~al.}(2017){Crossett}, {Pimbblet}, {Jones}, {Brown}, \&
  {Stott}}]{Crossett17}
{Crossett}, J.~P., {Pimbblet}, K.~A., {Jones}, D.~H., {Brown}, M.~J.~I., \&
  {Stott}, J.~P. 2017, \mnras, 464, 480

\bibitem[{{Croton} {et~al.}(2006){Croton}, {Springel}, {White}, {De Lucia},
  {Frenk}, {Gao}, {Jenkins}, {Kauffmann}, {Navarro}, \& {Yoshida}}]{Croton06}
{Croton}, D.~J., {Springel}, V., {White}, S.~D.~M., {et~al.} 2006, \mnras, 365,
  11

\bibitem[{{Darvish} {et~al.}(2017){Darvish}, {Mobasher}, {Martin}, {Sobral},
  {Scoville}, {Stroe}, {Hemmati}, \& {Kartaltepe}}]{Darvish17}
{Darvish}, B., {Mobasher}, B., {Martin}, D.~C., {et~al.} 2017, \apj, 837, 16

\bibitem[{{Darvish} {et~al.}(2015{\natexlab{a}}){Darvish}, {Mobasher},
  {Sobral}, {Hemmati}, {Nayyeri}, \& {Shivaei}}]{Darvish15b}
{Darvish}, B., {Mobasher}, B., {Sobral}, D., {et~al.} 2015{\natexlab{a}}, \apj,
  814, 84

\bibitem[{{Darvish} {et~al.}(2016){Darvish}, {Mobasher}, {Sobral}, {Rettura},
  {Scoville}, {Faisst}, \& {Capak}}]{Darvish16}
---. 2016, \apj, 825, 113

\bibitem[{{Darvish} {et~al.}(2015{\natexlab{b}}){Darvish}, {Mobasher},
  {Sobral}, {Scoville}, \& {Aragon-Calvo}}]{Darvish15a}
{Darvish}, B., {Mobasher}, B., {Sobral}, D., {Scoville}, N., \& {Aragon-Calvo},
  M. 2015{\natexlab{b}}, \apj, 805, 121

\bibitem[{{Darvish} {et~al.}(2014){Darvish}, {Sobral}, {Mobasher}, {Scoville},
  {Best}, {Sales}, \& {Smail}}]{Darvish14}
{Darvish}, B., {Sobral}, D., {Mobasher}, B., {et~al.} 2014, \apj, 796, 51

\bibitem[{{Davidzon} {et~al.}(2016){Davidzon}, {Cucciati}, {Bolzonella}, {De
  Lucia}, {Zamorani}, {Arnouts}, {Moutard}, {Ilbert}, {Garilli}, {Scodeggio},
  {Guzzo}, {Abbas}, {Adami}, {Bel}, {Bottini}, {Branchini}, {Cappi}, {Coupon},
  {de la Torre}, {Di Porto}, {Fritz}, {Franzetti}, {Fumana}, {Granett},
  {Guennou}, {Iovino}, {Krywult}, {Le Brun}, {Le F{\`e}vre}, {Maccagni},
  {Ma{\l}ek}, {Marulli}, {McCracken}, {Mellier}, {Moscardini}, {Polletta},
  {Pollo}, {Tasca}, {Tojeiro}, {Vergani}, \& {Zanichelli}}]{Davidzon16}
{Davidzon}, I., {Cucciati}, O., {Bolzonella}, M., {et~al.} 2016, \aap, 586, A23

\bibitem[{{De Lucia} {et~al.}(2006){De Lucia}, {Springel}, {White}, {Croton},
  \& {Kauffmann}}]{Delucia06}
{De Lucia}, G., {Springel}, V., {White}, S.~D.~M., {Croton}, D., \&
  {Kauffmann}, G. 2006, \mnras, 366, 499

\bibitem[{{Dekel} {et~al.}(2013){Dekel}, {Zolotov}, {Tweed}, {Cacciato},
  {Ceverino}, \& {Primack}}]{Dekel13}
{Dekel}, A., {Zolotov}, A., {Tweed}, D., {et~al.} 2013, \mnras, 435, 999

\bibitem[{{Duarte} \& {Mamon}(2014)}]{Duarte14}
{Duarte}, M., \& {Mamon}, G.~A. 2014, \mnras, 440, 1763

\bibitem[{{Duivenvoorden} {et~al.}(2016){Duivenvoorden}, {Oliver}, {Buat},
  {Darvish}, {Efstathiou}, {Farrah}, {Griffin}, {Hurley}, {Ibar}, {Jarvis},
  {Papadopoulos}, {Sargent}, {Scott}, {Scudder}, {Symeonidis}, {Vaccari},
  {Viero}, \& {Wang}}]{Duivenvoorden16}
{Duivenvoorden}, S., {Oliver}, S., {Buat}, V., {et~al.} 2016, \mnras, 462, 277

\bibitem[{{Efstathiou} \& {Moody}(2001)}]{Efstathiou01}
{Efstathiou}, G., \& {Moody}, S.~J. 2001, \mnras, 325, 1603

\bibitem[{{Ellison} {et~al.}(2013){Ellison}, {Mendel}, {Patton}, \&
  {Scudder}}]{Ellison13}
{Ellison}, S.~L., {Mendel}, J.~T., {Patton}, D.~R., \& {Scudder}, J.~M. 2013,
  \mnras, 435, 3627

\bibitem[{{Ellison} {et~al.}(2008){Ellison}, {Patton}, {Simard}, \&
  {McConnachie}}]{Ellison08}
{Ellison}, S.~L., {Patton}, D.~R., {Simard}, L., \& {McConnachie}, A.~W. 2008,
  \aj, 135, 1877

\bibitem[{{Ellison} {et~al.}(2010){Ellison}, {Patton}, {Simard}, {McConnachie},
  {Baldry}, \& {Mendel}}]{Ellison10}
{Ellison}, S.~L., {Patton}, D.~R., {Simard}, L., {et~al.} 2010, \mnras, 407,
  1514

\bibitem[{{Erfanianfar} {et~al.}(2016){Erfanianfar}, {Popesso}, {Finoguenov},
  {Wilman}, {Wuyts}, {Biviano}, {Salvato}, {Mirkazemi}, {Morselli}, {Ziparo},
  {Nandra}, {Lutz}, {Elbaz}, {Dickinson}, {Tanaka}, {Altieri}, {Aussel},
  {Bauer}, {Berta}, {Bielby}, {Brandt}, {Cappelluti}, {Cimatti}, {Cooper},
  {Fadda}, {Ilbert}, {Le Floch}, {Magnelli}, {Mulchaey}, {Nordon}, {Newman},
  {Poglitsch}, \& {Pozzi}}]{Erfanianfar16}
{Erfanianfar}, G., {Popesso}, P., {Finoguenov}, A., {et~al.} 2016, \mnras, 455,
  2839

\bibitem[{{Fabello} {et~al.}(2012){Fabello}, {Kauffmann}, {Catinella}, {Li},
  {Giovanelli}, \& {Haynes}}]{Fabello12}
{Fabello}, S., {Kauffmann}, G., {Catinella}, B., {et~al.} 2012, \mnras, 427,
  2841

\bibitem[{{Fabian}(2012)}]{Fabian12}
{Fabian}, A.~C. 2012, \araa, 50, 455

\bibitem[{{Fang} {et~al.}(2013){Fang}, {Faber}, {Koo}, \& {Dekel}}]{Fang13}
{Fang}, J.~J., {Faber}, S.~M., {Koo}, D.~C., \& {Dekel}, A. 2013, \apj, 776, 63

\bibitem[{{Finoguenov} {et~al.}(2007){Finoguenov}, {Guzzo}, {Hasinger},
  {Scoville}, {Aussel}, {B{\"o}hringer}, {Brusa}, {Capak}, {Cappelluti},
  {Comastri}, {Giodini}, {Griffiths}, {Impey}, {Koekemoer}, {Kneib},
  {Leauthaud}, {Le F{\`e}vre}, {Lilly}, {Mainieri}, {Massey}, {McCracken},
  {Mobasher}, {Murayama}, {Peacock}, {Sakelliou}, {Schinnerer}, {Silverman},
  {Smol{\v c}i{\'c}}, {Taniguchi}, {Tasca}, {Taylor}, {Trump}, \&
  {Zamorani}}]{Finoguenov07}
{Finoguenov}, A., {Guzzo}, L., {Hasinger}, G., {et~al.} 2007, \apjs, 172, 182

\bibitem[{{Fitzpatrick} \& {Graves}(2015)}]{Fitzpatrick15}
{Fitzpatrick}, P.~J., \& {Graves}, G.~J. 2015, \mnras, 447, 1383

\bibitem[{{Gaibler} {et~al.}(2012){Gaibler}, {Khochfar}, {Krause}, \&
  {Silk}}]{Gaibler12}
{Gaibler}, V., {Khochfar}, S., {Krause}, M., \& {Silk}, J. 2012, \mnras, 425,
  438

\bibitem[{{Gon{\c c}alves} {et~al.}(2012){Gon{\c c}alves}, {Martin},
  {Men{\'e}ndez-Delmestre}, {Wyder}, \& {Koekemoer}}]{goncalves12}
{Gon{\c c}alves}, T.~S., {Martin}, D.~C., {Men{\'e}ndez-Delmestre}, K.,
  {Wyder}, T.~K., \& {Koekemoer}, A. 2012, \apj, 759, 67

\bibitem[{{Guo} {et~al.}(2017){Guo}, {Bell}, {Lu}, {Koo}, {Faber}, {Koekemoer},
  {Kurczynski}, {Lee}, {Papovich}, {Chen}, {Dekel}, {Ferguson}, {Fontana},
  {Giavalisco}, {Kocevski}, {Nayyeri}, {P{\'e}rez-Gonz{\'a}lez}, {Pforr},
  {Rodr{\'{\i}}guez-Puebla}, \& {Santini}}]{Guo17}
{Guo}, Y., {Bell}, E.~F., {Lu}, Y., {et~al.} 2017, \apjl, 841, L22

\bibitem[{{G{\"u}rkan} {et~al.}(2015){G{\"u}rkan}, {Hardcastle}, {Jarvis},
  {Smith}, {Bourne}, {Dunne}, {Maddox}, {Ivison}, \& {Fritz}}]{Gurkan15}
{G{\"u}rkan}, G., {Hardcastle}, M.~J., {Jarvis}, M.~J., {et~al.} 2015, \mnras,
  452, 3776

\bibitem[{{Guzzo} {et~al.}(2007){Guzzo}, {Cassata}, {Finoguenov}, {Massey},
  {Scoville}, {Capak}, {Ellis}, {Mobasher}, {Taniguchi}, {Thompson}, {Ajiki},
  {Aussel}, {B{\"o}hringer}, {Brusa}, {Calzetti}, {Comastri}, {Franceschini},
  {Hasinger}, {Kasliwal}, {Kitzbichler}, {Kneib}, {Koekemoer}, {Leauthaud},
  {McCracken}, {Murayama}, {Nagao}, {Rhodes}, {Sanders}, {Sasaki}, {Shioya},
  {Tasca}, \& {Taylor}}]{Guzzo07}
{Guzzo}, L., {Cassata}, P., {Finoguenov}, A., {et~al.} 2007, \apjs, 172, 254

\bibitem[{{Haines} {et~al.}(2013){Haines}, {Pereira}, {Smith}, {Egami},
  {Sanderson}, {Babul}, {Finoguenov}, {Merluzzi}, {Busarello}, {Rawle}, \&
  {Okabe}}]{Haines13}
{Haines}, C.~P., {Pereira}, M.~J., {Smith}, G.~P., {et~al.} 2013, \apj, 775,
  126

\bibitem[{{Hatfield} \& {Jarvis}(2017)}]{Hatfield17}
{Hatfield}, P.~W., \& {Jarvis}, M.~J. 2017, \mnras, 472, 3570

\bibitem[{{Haynes} {et~al.}(2011){Haynes}, {Giovanelli}, {Martin}, {Hess},
  {Saintonge}, {Adams}, {Hallenbeck}, {Hoffman}, {Huang}, {Kent}, {Koopmann},
  {Papastergis}, {Stierwalt}, {Balonek}, {Craig}, {Higdon}, {Kornreich},
  {Miller}, {O'Donoghue}, {Olowin}, {Rosenberg}, {Spekkens}, {Troischt}, \&
  {Wilcots}}]{Haynes11}
{Haynes}, M.~P., {Giovanelli}, R., {Martin}, A.~M., {et~al.} 2011, \aj, 142,
  170

\bibitem[{{Henriques} {et~al.}(2017){Henriques}, {White}, {Thomas}, {Angulo},
  {Guo}, {Lemson}, \& {Wang}}]{Henriques17}
{Henriques}, B.~M.~B., {White}, S.~D.~M., {Thomas}, P.~A., {et~al.} 2017,
  \mnras, 469, 2626

\bibitem[{{Hopkins} \& {Elvis}(2010)}]{Hopkins10}
{Hopkins}, P.~F., \& {Elvis}, M. 2010, \mnras, 401, 7

\bibitem[{{Hopkins} {et~al.}(2014){Hopkins}, {Kere{\v s}}, {O{\~n}orbe},
  {Faucher-Gigu{\`e}re}, {Quataert}, {Murray}, \& {Bullock}}]{Hopkins14}
{Hopkins}, P.~F., {Kere{\v s}}, D., {O{\~n}orbe}, J., {et~al.} 2014, \mnras,
  445, 581

\bibitem[{{Huchra} \& {Geller}(1982)}]{Huchra82}
{Huchra}, J.~P., \& {Geller}, M.~J. 1982, \apj, 257, 423

\bibitem[{{Hung} {et~al.}(2016){Hung}, {Casey}, {Chiang}, {Capak}, {Cowley},
  {Darvish}, {Kacprzak}, {Kova{\v c}}, {Lilly}, {Nanayakkara}, {Spitler},
  {Tran}, \& {Yuan}}]{Hung16}
{Hung}, C.-L., {Casey}, C.~M., {Chiang}, Y.-K., {et~al.} 2016, \apj, 826, 130

\bibitem[{{Ilbert} {et~al.}(2009){Ilbert}, {Capak}, {Salvato}, {Aussel},
  {McCracken}, {Sanders}, {Scoville}, {Kartaltepe}, {Arnouts}, {Le Floc'h},
  {Mobasher}, {Taniguchi}, {Lamareille}, {Leauthaud}, {Sasaki}, {Thompson},
  {Zamojski}, {Zamorani}, {Bardelli}, {Bolzonella}, {Bongiorno}, {Brusa},
  {Caputi}, {Carollo}, {Contini}, {Cook}, {Coppa}, {Cucciati}, {de la Torre},
  {de Ravel}, {Franzetti}, {Garilli}, {Hasinger}, {Iovino}, {Kampczyk},
  {Kneib}, {Knobel}, {Kovac}, {Le Borgne}, {Le Brun}, {F{\`e}vre}, {Lilly},
  {Looper}, {Maier}, {Mainieri}, {Mellier}, {Mignoli}, {Murayama}, {Pell{\`o}},
  {Peng}, {P{\'e}rez-Montero}, {Renzini}, {Ricciardelli}, {Schiminovich},
  {Scodeggio}, {Shioya}, {Silverman}, {Surace}, {Tanaka}, {Tasca}, {Tresse},
  {Vergani}, \& {Zucca}}]{Ilbert09}
{Ilbert}, O., {Capak}, P., {Salvato}, M., {et~al.} 2009, \apj, 690, 1236

\bibitem[{{Kalfountzou} {et~al.}(2017){Kalfountzou}, {Stevens}, {Jarvis},
  {Hardcastle}, {Wilner}, {Elvis}, {Page}, {Trichas}, \&
  {Smith}}]{Kalfountzou17}
{Kalfountzou}, E., {Stevens}, J.~A., {Jarvis}, M.~J., {et~al.} 2017, \mnras,
  471, 28

\bibitem[{{Kauffmann} {et~al.}(2003){Kauffmann}, {Heckman}, {White}, {Charlot},
  {Tremonti}, {Brinchmann}, {Bruzual}, {Peng}, {Seibert}, {Bernardi},
  {Blanton}, {Brinkmann}, {Castander}, {Cs{\'a}bai}, {Fukugita}, {Ivezic},
  {Munn}, {Nichol}, {Padmanabhan}, {Thakar}, {Weinberg}, \&
  {York}}]{Kauffmann03}
{Kauffmann}, G., {Heckman}, T.~M., {White}, S.~D.~M., {et~al.} 2003, \mnras,
  341, 33

\bibitem[{{Kawinwanichakij} {et~al.}(2017){Kawinwanichakij}, {Papovich},
  {Quadri}, {Glazebrook}, {Kacprzak}, {Allen}, {Bell}, {Croton}, {Dekel},
  {Ferguson}, {Forrest}, {Grogin}, {Guo}, {Kocevski}, {Koekemoer}, {Labb{\'e}},
  {Lucas}, {Nanayakkara}, {Spitler}, {Straatman}, {Tran}, {Tomczak}, \& {van
  Dokkum}}]{Kawinwanichakij17}
{Kawinwanichakij}, L., {Papovich}, C., {Quadri}, R.~F., {et~al.} 2017, \apj,
  847, 134

\bibitem[{{Kewley} {et~al.}(2006){Kewley}, {Geller}, \& {Barton}}]{Kewley06}
{Kewley}, L.~J., {Geller}, M.~J., \& {Barton}, E.~J. 2006, \aj, 131, 2004

\bibitem[{{Kova{\v c}} {et~al.}(2014){Kova{\v c}}, {Lilly}, {Knobel},
  {Bschorr}, {Peng}, {Carollo}, {Contini}, {Kneib}, {Le F{\'e}vre}, {Mainieri},
  {Renzini}, {Scodeggio}, {Zamorani}, {Bardelli}, {Bolzonella}, {Bongiorno},
  {Caputi}, {Cucciati}, {de la Torre}, {de Ravel}, {Franzetti}, {Garilli},
  {Iovino}, {Kampczyk}, {Lamareille}, {Le Borgne}, {Le Brun}, {Maier},
  {Mignoli}, {Oesch}, {Pello}, {Montero}, {Presotto}, {Silverman}, {Tanaka},
  {Tasca}, {Tresse}, {Vergani}, {Zucca}, {Aussel}, {Koekemoer}, {Le Floc'h},
  {Moresco}, \& {Pozzetti}}]{Kovac14}
{Kova{\v c}}, K., {Lilly}, S.~J., {Knobel}, C., {et~al.} 2014, \mnras, 438, 717

\bibitem[{{Koyama} {et~al.}(2013){Koyama}, {Smail}, {Kurk}, {Geach}, {Sobral},
  {Kodama}, {Nakata}, {Swinbank}, {Best}, {Hayashi}, \& {Tadaki}}]{Koyama13a}
{Koyama}, Y., {Smail}, I., {Kurk}, J., {et~al.} 2013, \mnras, 434, 423

\bibitem[{{Laigle} {et~al.}(2016){Laigle}, {McCracken}, {Ilbert}, {Hsieh},
  {Davidzon}, {Capak}, {Hasinger}, {Silverman}, {Pichon}, {Coupon}, {Aussel},
  {Le Borgne}, {Caputi}, {Cassata}, {Chang}, {Civano}, {Dunlop}, {Fynbo},
  {Kartaltepe}, {Koekemoer}, {Le F{\`e}vre}, {Le Floc'h}, {Leauthaud}, {Lilly},
  {Lin}, {Marchesi}, {Milvang-Jensen}, {Salvato}, {Sanders}, {Scoville},
  {Smolcic}, {Stockmann}, {Taniguchi}, {Tasca}, {Toft}, {Vaccari}, \&
  {Zabl}}]{Laigle16}
{Laigle}, C., {McCracken}, H.~J., {Ilbert}, O., {et~al.} 2016, \apjs, 224, 24

\bibitem[{{Lee} {et~al.}(2015){Lee}, {Im}, {Kim}, {Lotz}, {McPartland}, {Peth},
  \& {Koekemoer}}]{Lee15}
{Lee}, S.-K., {Im}, M., {Kim}, J.-W., {et~al.} 2015, \apj, 810, 90

\bibitem[{{Lidman} {et~al.}(2012){Lidman}, {Suherli}, {Muzzin}, {Wilson},
  {Demarco}, {Brough}, {Rettura}, {Cox}, {DeGroot}, {Yee}, {Gilbank},
  {Hoekstra}, {Balogh}, {Ellingson}, {Hicks}, {Nantais}, {Noble}, {Lacy},
  {Surace}, \& {Webb}}]{Lidman12}
{Lidman}, C., {Suherli}, J., {Muzzin}, A., {et~al.} 2012, \mnras, 427, 550

\bibitem[{{Lin} {et~al.}(2010){Lin}, {Cooper}, {Jian}, {Koo}, {Patton}, {Yan},
  {Willmer}, {Coil}, {Chiueh}, {Croton}, {Gerke}, {Lotz}, {Guhathakurta}, \&
  {Newman}}]{Lin10}
{Lin}, L., {Cooper}, M.~C., {Jian}, H.-Y., {et~al.} 2010, \apj, 718, 1158

\bibitem[{{Martig} {et~al.}(2009){Martig}, {Bournaud}, {Teyssier}, \&
  {Dekel}}]{Martig09}
{Martig}, M., {Bournaud}, F., {Teyssier}, R., \& {Dekel}, A. 2009, \apj, 707,
  250

\bibitem[{{Martin} {et~al.}(2017){Martin}, {Gon{\c c}alves}, {Darvish},
  {Seibert}, \& {Schiminovich}}]{Martin17}
{Martin}, D.~C., {Gon{\c c}alves}, T.~S., {Darvish}, B., {Seibert}, M., \&
  {Schiminovich}, D. 2017, \apj, 842, 20

\bibitem[{{Martin} {et~al.}(2005){Martin}, {Fanson}, {Schiminovich},
  {Morrissey}, {Friedman}, {Barlow}, {Conrow}, {Grange}, {Jelinsky},
  {Milliard}, {Siegmund}, {Bianchi}, {Byun}, {Donas}, {Forster}, {Heckman},
  {Lee}, {Madore}, {Malina}, {Neff}, {Rich}, {Small}, {Surber}, {Szalay},
  {Welsh}, \& {Wyder}}]{Martin05}
{Martin}, D.~C., {Fanson}, J., {Schiminovich}, D., {et~al.} 2005, \apjl, 619,
  L1

\bibitem[{{Martin} {et~al.}(2007){Martin}, {Wyder}, {Schiminovich}, {Barlow},
  {Forster}, {Friedman}, {Morrissey}, {Neff}, {Seibert}, {Small}, {Welsh},
  {Bianchi}, {Donas}, {Heckman}, {Lee}, {Madore}, {Milliard}, {Rich}, {Szalay},
  \& {Yi}}]{Martin07}
{Martin}, D.~C., {Wyder}, T.~K., {Schiminovich}, D., {et~al.} 2007, \apjs, 173,
  342

\bibitem[{{McGee} {et~al.}(2014){McGee}, {Bower}, \& {Balogh}}]{McGee14}
{McGee}, S.~L., {Bower}, R.~G., \& {Balogh}, M.~L. 2014, \mnras, 442, L105

\bibitem[{{McIntosh} {et~al.}(2008){McIntosh}, {Guo}, {Hertzberg}, {Katz},
  {Mo}, {van den Bosch}, \& {Yang}}]{McIntosh08}
{McIntosh}, D.~H., {Guo}, Y., {Hertzberg}, J., {et~al.} 2008, \mnras, 388, 1537

\bibitem[{{Mihos} \& {Hernquist}(1996)}]{Mihos96}
{Mihos}, J.~C., \& {Hernquist}, L. 1996, \apj, 464, 641

\bibitem[{{Mihos} {et~al.}(1992){Mihos}, {Richstone}, \& {Bothun}}]{Mihos92}
{Mihos}, J.~C., {Richstone}, D.~O., \& {Bothun}, G.~D. 1992, \apj, 400, 153

\bibitem[{{Muldrew} {et~al.}(2012){Muldrew}, {Croton}, {Skibba}, {Pearce},
  {Ann}, {Baldry}, {Brough}, {Choi}, {Conselice}, {Cowan}, {Gallazzi}, {Gray},
  {Gr{\"u}tzbauch}, {Li}, {Park}, {Pilipenko}, {Podgorzec}, {Robotham},
  {Wilman}, {Yang}, {Zhang}, \& {Zibetti}}]{Muldrew12}
{Muldrew}, S.~I., {Croton}, D.~J., {Skibba}, R.~A., {et~al.} 2012, \mnras, 419,
  2670

\bibitem[{{Nantais} {et~al.}(2017){Nantais}, {Muzzin}, {van der Burg},
  {Wilson}, {Lidman}, {Foltz}, {DeGroot}, {Noble}, {Cooper}, \&
  {Demarco}}]{Nantais17}
{Nantais}, J.~B., {Muzzin}, A., {van der Burg}, R.~F.~J., {et~al.} 2017,
  \mnras, 465, L104

\bibitem[{{Noeske} {et~al.}(2007){Noeske}, {Weiner}, {Faber}, {Papovich},
  {Koo}, {Somerville}, {Bundy}, {Conselice}, {Newman}, {Schiminovich}, {Le
  Floc'h}, {Coil}, {Rieke}, {Lotz}, {Primack}, {Barmby}, {Cooper}, {Davis},
  {Ellis}, {Fazio}, {Guhathakurta}, {Huang}, {Kassin}, {Martin}, {Phillips},
  {Rich}, {Small}, {Willmer}, \& {Wilson}}]{Noeske07a}
{Noeske}, K.~G., {Weiner}, B.~J., {Faber}, S.~M., {et~al.} 2007, \apjl, 660,
  L43

\bibitem[{{Nogueira-Cavalcante} {et~al.}(2018){Nogueira-Cavalcante}, {Gon{\c
  c}alves}, {Men{\'e}ndez-Delmestre}, \& {Sheth}}]{Cavalcante18}
{Nogueira-Cavalcante}, J.~P., {Gon{\c c}alves}, T.~S.,
  {Men{\'e}ndez-Delmestre}, K., \& {Sheth}, K. 2018, \mnras, 473, 1346

\bibitem[{{Obreschkow} \& {Rawlings}(2009)}]{Obreschkow09}
{Obreschkow}, D., \& {Rawlings}, S. 2009, \mnras, 394, 1857

\bibitem[{{Pasquali} {et~al.}(2010){Pasquali}, {Gallazzi}, {Fontanot}, {van den
  Bosch}, {De Lucia}, {Mo}, \& {Yang}}]{Pasquali10}
{Pasquali}, A., {Gallazzi}, A., {Fontanot}, F., {et~al.} 2010, \mnras, 407, 937

\bibitem[{{Patel} {et~al.}(2009){Patel}, {Holden}, {Kelson}, {Illingworth}, \&
  {Franx}}]{Patel09}
{Patel}, S.~G., {Holden}, B.~P., {Kelson}, D.~D., {Illingworth}, G.~D., \&
  {Franx}, M. 2009, \apjl, 705, L67

\bibitem[{{Patel} {et~al.}(2011){Patel}, {Kelson}, {Holden}, {Franx}, \&
  {Illingworth}}]{Patel11}
{Patel}, S.~G., {Kelson}, D.~D., {Holden}, B.~P., {Franx}, M., \&
  {Illingworth}, G.~D. 2011, \apj, 735, 53

\bibitem[{{Patton} \& {Atfield}(2008)}]{Patton08}
{Patton}, D.~R., \& {Atfield}, J.~E. 2008, \apj, 685, 235

\bibitem[{{Peng} {et~al.}(2015){Peng}, {Maiolino}, \& {Cochrane}}]{Peng15}
{Peng}, Y., {Maiolino}, R., \& {Cochrane}, R. 2015, \nat, 521, 192

\bibitem[{{Peng} {et~al.}(2012){Peng}, {Lilly}, {Renzini}, \&
  {Carollo}}]{Peng12}
{Peng}, Y.-j., {Lilly}, S.~J., {Renzini}, A., \& {Carollo}, M. 2012, \apj, 757,
  4

\bibitem[{{Peng} {et~al.}(2010){Peng}, {Lilly}, {Kova{\v c}}, {Bolzonella},
  {Pozzetti}, {Renzini}, {Zamorani}, {Ilbert}, {Knobel}, {Iovino}, {Maier},
  {Cucciati}, {Tasca}, {Carollo}, {Silverman}, {Kampczyk}, {de Ravel},
  {Sanders}, {Scoville}, {Contini}, {Mainieri}, {Scodeggio}, {Kneib}, {Le
  F{\`e}vre}, {Bardelli}, {Bongiorno}, {Caputi}, {Coppa}, {de la Torre},
  {Franzetti}, {Garilli}, {Lamareille}, {Le Borgne}, {Le Brun}, {Mignoli},
  {Perez Montero}, {Pello}, {Ricciardelli}, {Tanaka}, {Tresse}, {Vergani},
  {Welikala}, {Zucca}, {Oesch}, {Abbas}, {Barnes}, {Bordoloi}, {Bottini},
  {Cappi}, {Cassata}, {Cimatti}, {Fumana}, {Hasinger}, {Koekemoer},
  {Leauthaud}, {Maccagni}, {Marinoni}, {McCracken}, {Memeo}, {Meneux}, {Nair},
  {Porciani}, {Presotto}, \& {Scaramella}}]{Peng10}
{Peng}, Y.-j., {Lilly}, S.~J., {Kova{\v c}}, K., {et~al.} 2010, \apj, 721, 193

\bibitem[{{Perez} {et~al.}(2009){Perez}, {Tissera}, {Padilla}, {Alonso}, \&
  {Lambas}}]{Perez09}
{Perez}, J., {Tissera}, P., {Padilla}, N., {Alonso}, M.~S., \& {Lambas}, D.~G.
  2009, \mnras, 399, 1157

\bibitem[{{Poggianti} {et~al.}(2016){Poggianti}, {Fasano}, {Omizzolo},
  {Gullieuszik}, {Bettoni}, {Moretti}, {Paccagnella}, {Jaff{\'e}}, {Vulcani},
  {Fritz}, {Couch}, \& {D'Onofrio}}]{Poggianti16}
{Poggianti}, B.~M., {Fasano}, G., {Omizzolo}, A., {et~al.} 2016, \aj, 151, 78

\bibitem[{{Poggianti} {et~al.}(2017){Poggianti}, {Jaff{\'e}}, {Moretti},
  {Gullieuszik}, {Radovich}, {Tonnesen}, {Fritz}, {Bettoni}, {Vulcani},
  {Fasano}, {Bellhouse}, {Hau}, \& {Omizzolo}}]{Poggianti17}
{Poggianti}, B.~M., {Jaff{\'e}}, Y.~L., {Moretti}, A., {et~al.} 2017, \nat,
  548, 304

\bibitem[{{Pozzetti} {et~al.}(2010){Pozzetti}, {Bolzonella}, {Zucca},
  {Zamorani}, {Lilly}, {Renzini}, {Moresco}, {Mignoli}, {Cassata}, {Tasca},
  {Lamareille}, {Maier}, {Meneux}, {Halliday}, {Oesch}, {Vergani}, {Caputi},
  {Kova{\v c}}, {Cimatti}, {Cucciati}, {Iovino}, {Peng}, {Carollo}, {Contini},
  {Kneib}, {Le F{\'e}vre}, {Mainieri}, {Scodeggio}, {Bardelli}, {Bongiorno},
  {Coppa}, {de la Torre}, {de Ravel}, {Franzetti}, {Garilli}, {Kampczyk},
  {Knobel}, {Le Borgne}, {Le Brun}, {Pell{\`o}}, {Perez Montero},
  {Ricciardelli}, {Silverman}, {Tanaka}, {Tresse}, {Abbas}, {Bottini}, {Cappi},
  {Guzzo}, {Koekemoer}, {Leauthaud}, {Maccagni}, {Marinoni}, {McCracken},
  {Memeo}, {Porciani}, {Scaramella}, {Scarlata}, \& {Scoville}}]{Pozzetti10}
{Pozzetti}, L., {Bolzonella}, M., {Zucca}, E., {et~al.} 2010, \aap, 523, A13

\bibitem[{{Quadri} {et~al.}(2012){Quadri}, {Williams}, {Franx}, \&
  {Hildebrandt}}]{Quadri12}
{Quadri}, R.~F., {Williams}, R.~J., {Franx}, M., \& {Hildebrandt}, H. 2012,
  \apj, 744, 88

\bibitem[{{Robotham} {et~al.}(2014){Robotham}, {Driver}, {Davies}, {Hopkins},
  {Baldry}, {Agius}, {Bauer}, {Bland-Hawthorn}, {Brough}, {Brown}, {Cluver},
  {De Propris}, {Drinkwater}, {Holwerda}, {Kelvin}, {Lara-Lopez}, {Liske},
  {L{\'o}pez-S{\'a}nchez}, {Loveday}, {Mahajan}, {McNaught-Roberts}, {Moffett},
  {Norberg}, {Obreschkow}, {Owers}, {Penny}, {Pimbblet}, {Prescott}, {Taylor},
  {van Kampen}, \& {Wilkins}}]{Robotham14}
{Robotham}, A.~S.~G., {Driver}, S.~P., {Davies}, L.~J.~M., {et~al.} 2014,
  \mnras, 444, 3986

\bibitem[{{Rowlands} {et~al.}(2018){Rowlands}, {Wild}, {Bourne}, {Bremer},
  {Brough}, {Driver}, {Hopkins}, {Owers}, {Phillipps}, {Pimbblet}, {Sansom},
  {Wang}, {Alpaslan}, {Bland-Hawthorn}, {Colless}, {Holwerda}, \&
  {Taylor}}]{Rowlands18}
{Rowlands}, K., {Wild}, V., {Bourne}, N., {et~al.} 2018, \mnras, 473, 1168

\bibitem[{{Salpeter}(1955)}]{Salpeter55}
{Salpeter}, E.~E. 1955, \apj, 121, 161

\bibitem[{{Scoville} {et~al.}(2007){Scoville}, {Aussel}, {Brusa}, {Capak},
  {Carollo}, {Elvis}, {Giavalisco}, {Guzzo}, {Hasinger}, {Impey}, {Kneib},
  {LeFevre}, {Lilly}, {Mobasher}, {Renzini}, {Rich}, {Sanders}, {Schinnerer},
  {Schminovich}, {Shopbell}, {Taniguchi}, \& {Tyson}}]{Scoville07}
{Scoville}, N., {Aussel}, H., {Brusa}, M., {et~al.} 2007, \apjs, 172, 1

\bibitem[{{Scoville} {et~al.}(2013){Scoville}, {Arnouts}, {Aussel}, {Benson},
  {Bongiorno}, {Bundy}, {Calvo}, {Capak}, {Carollo}, {Civano}, {Dunlop},
  {Elvis}, {Faisst}, {Finoguenov}, {Fu}, {Giavalisco}, {Guo}, {Ilbert},
  {Iovino}, {Kajisawa}, {Kartaltepe}, {Leauthaud}, {Le F{\`e}vre}, {LeFloch},
  {Lilly}, {Liu}, {Manohar}, {Massey}, {Masters}, {McCracken}, {Mobasher},
  {Peng}, {Renzini}, {Rhodes}, {Salvato}, {Sanders}, {Sarvestani}, {Scarlata},
  {Schinnerer}, {Sheth}, {Shopbell}, {Smol{\v c}i{\'c}}, {Taniguchi}, {Taylor},
  {White}, \& {Yan}}]{Scoville13}
{Scoville}, N., {Arnouts}, S., {Aussel}, H., {et~al.} 2013, \apjs, 206, 3

\bibitem[{{Seibert} {et~al.}(2012){Seibert}, {Wyder}, {Neill}, {Madore},
  {Bianchi}, {Smith}, {Shiao}, {Schiminovich}, {Rich}, {Conrow}, {Martin}, \&
  {GALEX Catalog Team}}]{Seibert12}
{Seibert}, M., {Wyder}, T., {Neill}, J., {et~al.} 2012, in American
  Astronomical Society Meeting Abstracts, Vol. 219, American Astronomical
  Society Meeting Abstracts \#219, 340.01

\bibitem[{{Shankar} {et~al.}(2015){Shankar}, {Buchan}, {Rettura}, {Bouillot},
  {Moreno}, {Licitra}, {Bernardi}, {Huertas-Company}, {Mei}, {Ascaso}, {Sheth},
  {Delaye}, \& {Raichoor}}]{Shankar15}
{Shankar}, F., {Buchan}, S., {Rettura}, A., {et~al.} 2015, \apj, 802, 73

\bibitem[{{Sheth} {et~al.}(2005){Sheth}, {Vogel}, {Regan}, {Thornley}, \&
  {Teuben}}]{Sheth05}
{Sheth}, K., {Vogel}, S.~N., {Regan}, M.~W., {Thornley}, M.~D., \& {Teuben},
  P.~J. 2005, \apj, 632, 217

\bibitem[{{Shivaei} {et~al.}(2015){Shivaei}, {Reddy}, {Shapley}, {Kriek},
  {Siana}, {Mobasher}, {Coil}, {Freeman}, {Sanders}, {Price}, {de Groot}, \&
  {Azadi}}]{Shivaei15b}
{Shivaei}, I., {Reddy}, N.~A., {Shapley}, A.~E., {et~al.} 2015, \apj, 815, 98

\bibitem[{{Silk} \& {Nusser}(2010)}]{Silk10}
{Silk}, J., \& {Nusser}, A. 2010, \apj, 725, 556

\bibitem[{{Smethurst} {et~al.}(2017){Smethurst}, {Lintott}, {Bamford}, {Hart},
  {Kruk}, {Masters}, {Nichol}, \& {Simmons}}]{Smethurst17}
{Smethurst}, R.~J., {Lintott}, C.~J., {Bamford}, S.~P., {et~al.} 2017, \mnras,
  469, 3670

\bibitem[{{Sobral} {et~al.}(2011){Sobral}, {Best}, {Smail}, {Geach},
  {Cirasuolo}, {Garn}, \& {Dalton}}]{Sobral11}
{Sobral}, D., {Best}, P.~N., {Smail}, I., {et~al.} 2011, \mnras, 411, 675

\bibitem[{{Sobral} {et~al.}(2015){Sobral}, {Stroe}, {Dawson}, {Wittman}, {Jee},
  {R{\"o}ttgering}, {van Weeren}, \& {Br{\"u}ggen}}]{Sobral15}
{Sobral}, D., {Stroe}, A., {Dawson}, W.~A., {et~al.} 2015, \mnras, 450, 630

\bibitem[{{Somerville} {et~al.}(2008){Somerville}, {Hopkins}, {Cox},
  {Robertson}, \& {Hernquist}}]{Somerville08}
{Somerville}, R.~S., {Hopkins}, P.~F., {Cox}, T.~J., {Robertson}, B.~E., \&
  {Hernquist}, L. 2008, \mnras, 391, 481

\bibitem[{{Speagle} {et~al.}(2014){Speagle}, {Steinhardt}, {Capak}, \&
  {Silverman}}]{Speagle14}
{Speagle}, J.~S., {Steinhardt}, C.~L., {Capak}, P.~L., \& {Silverman}, J.~D.
  2014, \apjs, 214, 15

\bibitem[{{Springel} {et~al.}(2005){Springel}, {White}, {Jenkins}, {Frenk},
  {Yoshida}, {Gao}, {Navarro}, {Thacker}, {Croton}, {Helly}, {Peacock}, {Cole},
  {Thomas}, {Couchman}, {Evrard}, {Colberg}, \& {Pearce}}]{Springel05}
{Springel}, V., {White}, S.~D.~M., {Jenkins}, A., {et~al.} 2005, \nat, 435, 629

\bibitem[{{Strauss} {et~al.}(2002){Strauss}, {Weinberg}, {Lupton}, {Narayanan},
  {Annis}, {Bernardi}, {Blanton}, {Burles}, {Connolly}, {Dalcanton}, {Doi},
  {Eisenstein}, {Frieman}, {Fukugita}, {Gunn}, {Ivezi{\'c}}, {Kent}, {Kim},
  {Knapp}, {Kron}, {Munn}, {Newberg}, {Nichol}, {Okamura}, {Quinn}, {Richmond},
  {Schlegel}, {Shimasaku}, {SubbaRao}, {Szalay}, {Vanden Berk}, {Vogeley},
  {Yanny}, {Yasuda}, {York}, \& {Zehavi}}]{Strauss02}
{Strauss}, M.~A., {Weinberg}, D.~H., {Lupton}, R.~H., {et~al.} 2002, \aj, 124,
  1810

\bibitem[{{Stroe} {et~al.}(2015){Stroe}, {Sobral}, {Dawson}, {Jee}, {Hoekstra},
  {Wittman}, {van Weeren}, {Br{\"u}ggen}, \& {R{\"o}ttgering}}]{Stroe15}
{Stroe}, A., {Sobral}, D., {Dawson}, W., {et~al.} 2015, \mnras, 450, 646

\bibitem[{{Tinker} \& {Wetzel}(2010)}]{Tinker10}
{Tinker}, J.~L., \& {Wetzel}, A.~R. 2010, \apj, 719, 88

\bibitem[{{Tonnesen} \& {Cen}(2012)}]{Tonnesen12}
{Tonnesen}, S., \& {Cen}, R. 2012, \mnras, 425, 2313

\bibitem[{{van de Voort} {et~al.}(2017){van de Voort}, {Bah{\'e}}, {Bower},
  {Correa}, {Crain}, {Schaye}, \& {Theuns}}]{vandeVoort17}
{van de Voort}, F., {Bah{\'e}}, Y.~M., {Bower}, R.~G., {et~al.} 2017, \mnras,
  466, 3460

\bibitem[{{van der Wel} {et~al.}(2016){van der Wel}, {Noeske}, {Bezanson},
  {Pacifici}, {Gallazzi}, {Franx}, {Mu{\~n}oz-Mateos}, {Bell}, {Brammer},
  {Charlot}, {Chauk{\'e}}, {Labb{\'e}}, {Maseda}, {Muzzin}, {Rix}, {Sobral},
  {van de Sande}, {van Dokkum}, {Wild}, \& {Wolf}}]{vanderwel16}
{van der Wel}, A., {Noeske}, K., {Bezanson}, R., {et~al.} 2016, \apjs, 223, 29

\bibitem[{{Vulcani} {et~al.}(2010){Vulcani}, {Poggianti}, {Finn}, {Rudnick},
  {Desai}, \& {Bamford}}]{Vulcani10}
{Vulcani}, B., {Poggianti}, B.~M., {Finn}, R.~A., {et~al.} 2010, \apjl, 710, L1

\bibitem[{{Wagner} {et~al.}(2012){Wagner}, {Bicknell}, \& {Umemura}}]{Wagner12}
{Wagner}, A.~Y., {Bicknell}, G.~V., \& {Umemura}, M. 2012, \apj, 757, 136

\bibitem[{{Wetzel} {et~al.}(2014){Wetzel}, {Tinker}, {Conroy}, \& {van den
  Bosch}}]{Wetzel14}
{Wetzel}, A.~R., {Tinker}, J.~L., {Conroy}, C., \& {van den Bosch}, F.~C. 2014,
  \mnras, 439, 2687

\bibitem[{{Whitaker} {et~al.}(2012){Whitaker}, {van Dokkum}, {Brammer}, \&
  {Franx}}]{Whitaker12}
{Whitaker}, K.~E., {van Dokkum}, P.~G., {Brammer}, G., \& {Franx}, M. 2012,
  \apjl, 754, L29

\bibitem[{{Wuyts} {et~al.}(2011){Wuyts}, {F{\"o}rster Schreiber}, {van der
  Wel}, {Magnelli}, {Guo}, {Genzel}, {Lutz}, {Aussel}, {Barro}, {Berta},
  {Cava}, {Graci{\'a}-Carpio}, {Hathi}, {Huang}, {Kocevski}, {Koekemoer},
  {Lee}, {Le Floc'h}, {McGrath}, {Nordon}, {Popesso}, {Pozzi}, {Riguccini},
  {Rodighiero}, {Saintonge}, \& {Tacconi}}]{Wuyts11}
{Wuyts}, S., {F{\"o}rster Schreiber}, N.~M., {van der Wel}, A., {et~al.} 2011,
  \apj, 742, 96

\bibitem[{{Xu} {et~al.}(2012){Xu}, {Zhao}, {Scoville}, {Capak}, {Drory}, \&
  {Gao}}]{Xu12}
{Xu}, C.~K., {Zhao}, Y., {Scoville}, N., {et~al.} 2012, \apj, 747, 85

\bibitem[{{Zamojski} {et~al.}(2007){Zamojski}, {Schiminovich}, {Rich},
  {Mobasher}, {Koekemoer}, {Capak}, {Taniguchi}, {Sasaki}, {McCracken},
  {Mellier}, {Bertin}, {Aussel}, {Sanders}, {Le F{\`e}vre}, {Ilbert},
  {Salvato}, {Thompson}, {Kartaltepe}, {Scoville}, {Barlow}, {Forster},
  {Friedman}, {Martin}, {Morrissey}, {Neff}, {Seibert}, {Small}, {Wyder},
  {Bianchi}, {Donas}, {Heckman}, {Lee}, {Madore}, {Milliard}, {Szalay},
  {Welsh}, \& {Yi}}]{Zamojski07}
{Zamojski}, M.~A., {Schiminovich}, D., {Rich}, R.~M., {et~al.} 2007, \apjs,
  172, 468

\end{thebibliography}

\end{document}